\documentclass[amsmath,amssymb,aps,prl,tightenlines,twocolumn]{revtex4-1}

\usepackage[english]{babel}
\usepackage{afterpage}
\usepackage{amsmath}
\usepackage{graphicx}
\usepackage{dcolumn}
\usepackage{bm}
\usepackage{color}
\usepackage{slashed}
\usepackage{latexsym}
\usepackage{url}
\usepackage{ulem}
\usepackage[version=4]{mhchem}

\begin{document}

\title{Anomalous photo-induced band renormalization in correlated materials: Case study of Ta$_2$NiSe$_5$}

\author{Lei Geng$^1$}
\author{Xiulan Liu$^1$}
\author{Jianing Zhang$^1$}
\author{Denis Gole{\v{z}}$^{2,3}$}
\email{denis.golez@ijs.si}
\author{Liang-You Peng$^{1,4,5,6}$}
\email{liangyou.peng@pku.edu.cn}

\affiliation{
$^1$State Key Laboratory for Mesoscopic Physics and Frontiers Science Center for Nano-optoelectronics, School of Physics, Peking University, 100871 Beijing, China\\
$^2$Jozef Stefan Institute, Jamova 39, SI-1000 Ljubljana, Slovenia \\
$^3$Faculty of Mathematics and Physics, University of Ljubljana, Jadranska 19, SI-1000 Ljubljana, Slovenia \\
$^4$Collaborative Innovation Center of Quantum Matter, Beijing 100871, China\\
$^5$Collaborative Innovation Center of Extreme Optics, Shanxi University, 030006 Taiyuan, China \\
$^6$Beijing Academy of Quantum Information Sciences, Beijing 100193, China
}

\begin{abstract}
 We investigate the anomalous photo-induced band renormalization in correlated materials, exemplified by the case of Ta$_2$NiSe$_5$. The manifestation of this anomaly is characterized by the alternating direction of band shift in response to changes in the laser parameters or electron momentum. We attribute the phenomena to the band inversion of the material and the selective excitation of a high-lying flat band, leading to the competition between the Hartree shift and the order collapse. These findings are based on {\it ab initio} determined effective model for Ta$_2$NiSe$_5$, in which we incorporate high-lying states and the time-dependent GW simulation to follow the non-equilibrium dynamics induced by the laser. Our findings reveal the sensitivity of the non-equilibrium electronic dynamics to the band structure and laser protocols, providing valuable guidance for the selection of suitable materials and lasers in the engineering of band structures.
\end{abstract}
\maketitle
The development of laser sources has provided an unprecedented opportunity to manipulate the properties of solid-state materials by strong light pulses~\cite{oka2019floquet,de2021colloquium,aoki2014nonequilibrium,murakami2023photo}. Ultrafast lasers can induce long-lived non-equilibrium states in quantum materials, showcasing distinctive features not observed in their equilibrium counterparts. Recent experiments have demonstrated the potential of photo-induced non-equilibrium states in modifying ferromagnetism~\cite{disa2023photo}, superconductivity~\cite{mankowsky2014nonlinear,mitrano2016possible,Fausti11} or charge-density waves~\cite{Stojchevska2014,Vaskivskyi2016,kogar2020,Zong2019a}. As material's properties are closely related to its electronic band structure, it is important to understand how it can be efficiently manipulated. The understanding is further strenghten by remarkable development of time- and angle-resolved photoemission spectroscopy~(ARPES) which provides a direct insight in the light induced bandgap renormalization~\cite{boschini2023time,Perfetti2006,Perfetti2008,Ligges2018,mor2017ultrafast,zhou2023pseudospin,liu2019direct,wegkamp2014instantaneous,baldini2023spontaneous,andreatta2019,puppin2022}.

The excitonic insulator (EI) is one of the most elusive states of matter. Despite its conceptual proposal decades ago~\cite{mott1961transition,jerome1967excitonic,halperin1968possible} and the implementation in artificial systems~\cite{eisenstein2014exciton,li2017excitonic,liu2017quantum}, providing conclusive evidence for its presence in real materials remains a significant challenge. The two most extensively studied candidates for the EI are 1T-\ce{TiSe2}~\cite{cercellier2007evidence,monney2010,kogar2017} and \ce{Ta2NiSe5} (TNS)~\cite{Wakisaka2009,kaneko2013orthorhombic,sugimoto2018strong,seki2014,Lu2017,Kim2021,Volkov2021,ye2021,Guan2023}. For TNS, ARPES results indeed showed the presence of the flat valence band top below a certain temperature, indicative of an ordered phase. Its origin has recently been highly debated. Valuable insights were given by time-resolved ARPES after high frequency excitations. The most important observation is the transient downward shift of the valence band, primarily localized near the $\Gamma$ point, distinguishing the response from the typical band renormalization in semiconductors~\cite{mor2017ultrafast,murakami2017photoinduced,golevz2022unveiling,baldini2023spontaneous,saha2021photoinduced}. In addition, this phenomenon is observed only above a certain pump intensity~\cite{mor2017ultrafast}. However, other experiments utilizing a different pump laser frequency reported a monotonic upward shift of the valence band~\cite{tang2020non}. These diverse results show that the direction of band shift in TNS is not only dependent on the electron momentum but also on laser intensities and frequencies, which provide compelling examples for the study of the anomalous band renormalization.

In this work, we combine the {\it ab-initio} determined band structure of TNS~\cite{mazza2020nature} with the self-consistent time-dependent GW approximation at finite temperature~\cite{golevz2016photoinduced,golevz2022unveiling,schuler2020nessi,keldysh1965diagram,kadanoff1962quantum} to study the equilibrium and non-equilibrium properties of TNS. Our results can consistently reproduce multiple phenomena observed in experiments. The underlying causes of the anomalous band renormalization are analyzed. A set of higher-lying flat orbitals not only play a role in symmetry breaking for equilibrium but also are responsible for a long-lived nonthermal state under a resonant excitation with respect to the valence band. The nonthermal population of the flat band significantly influences the competition between the Hartree shift and the order collapse, which can lead either to the downward or upward band shift depending on the pump parameters. The momentum dependence of the band renormalization is attributed to the band inversion of TNS. The investigation into anomalous band renormalization in TNS also provides insights for non-equilibrium processes in other materials.

\begin{figure}
   \includegraphics[width=\linewidth]{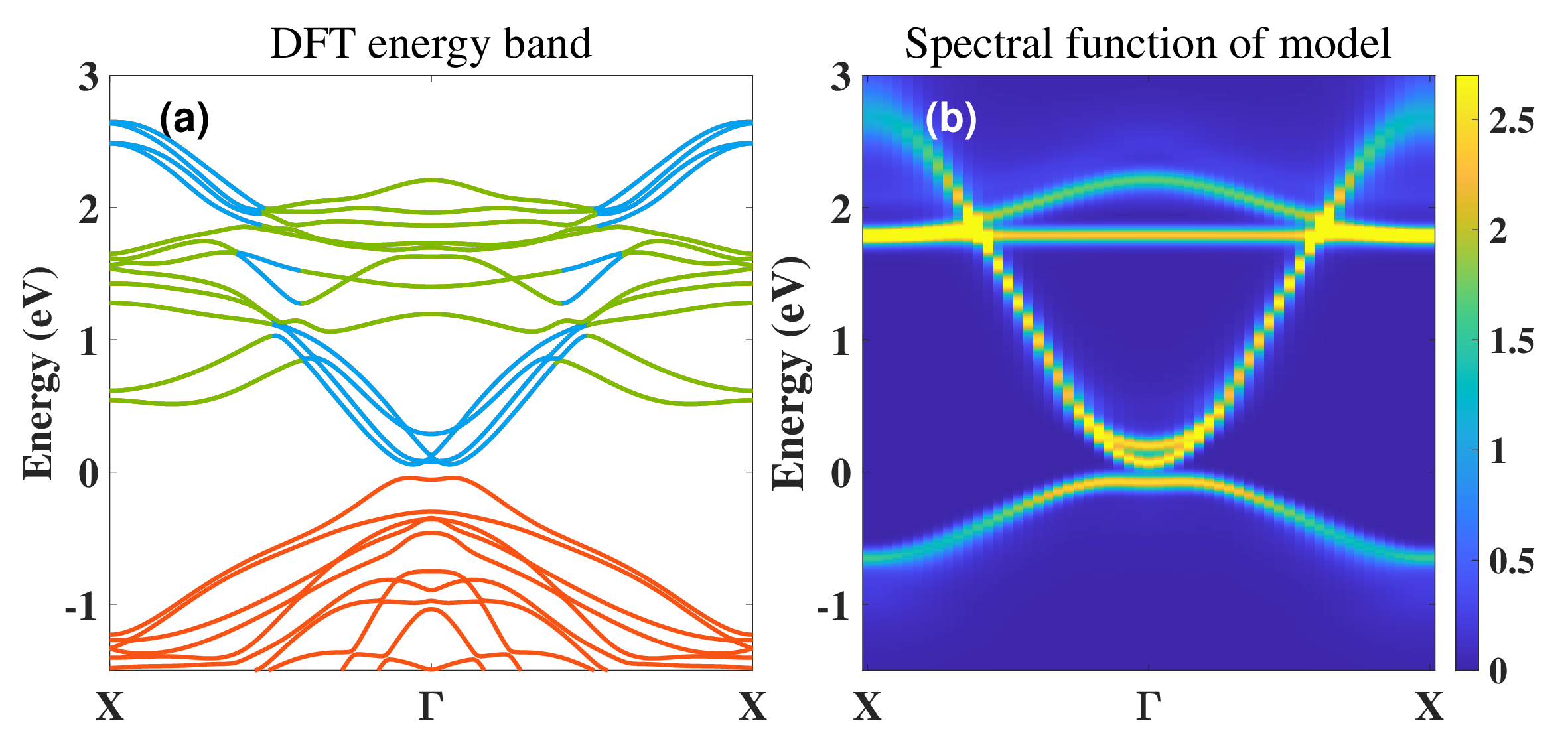}\quad
   \includegraphics[width=\linewidth]{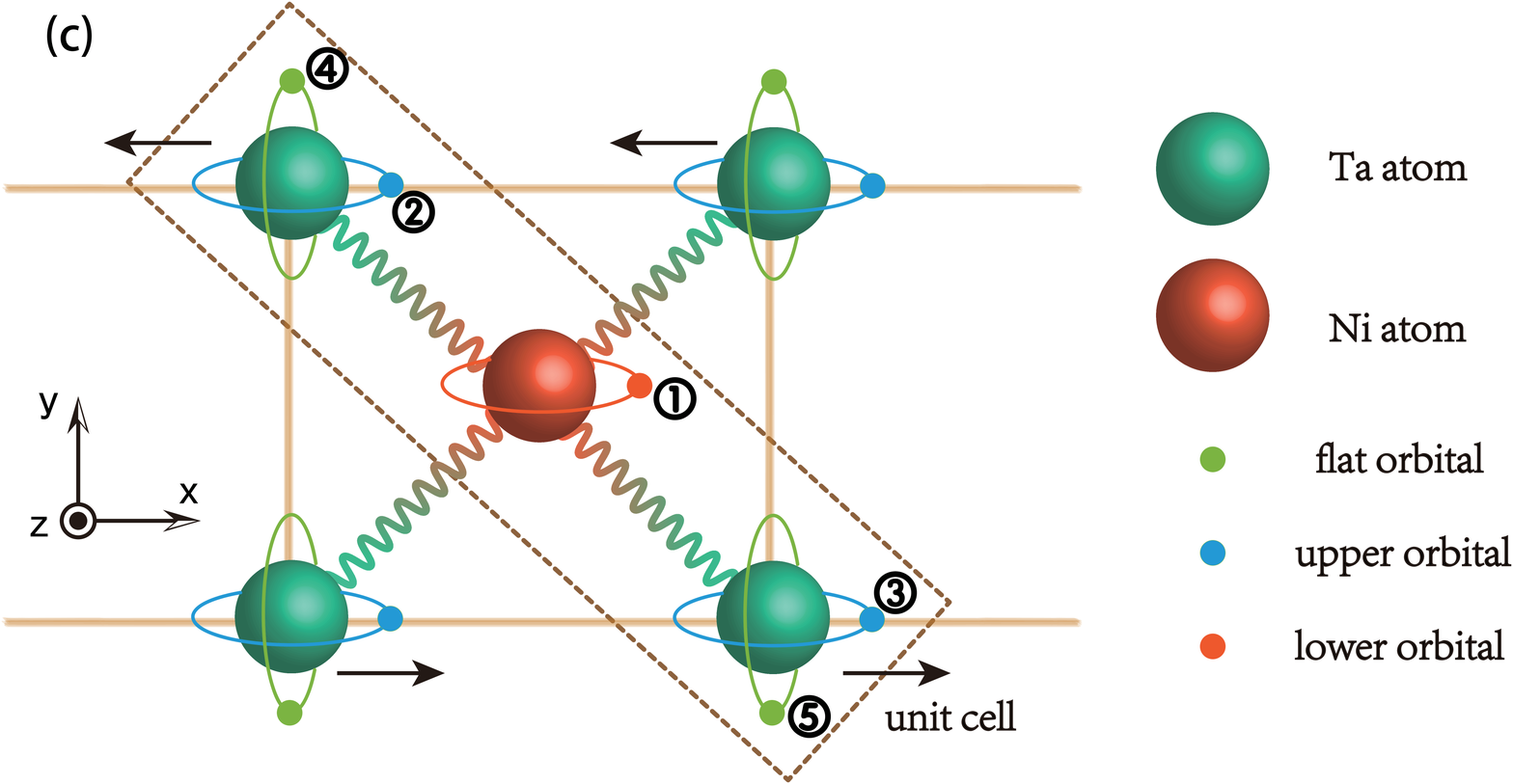}
   \caption{(a) The energy bands of the TNS in the excitonic phase from the DFT calculation. The bands in orange, blue and green color represent the lower, upper and flat bands respectively. (b) The spectral function of the model in the equilibrium excitonic phase with the temperature $T=116~{\rm K}$. (c) The sketch for the minimal model. There are three atoms and five orbitals in a unit cell. Orbital 1 on the \ce{Ni} atom is the lower orbital. Orbitals 2 and 3 on two \ce{Ta} atoms are upper orbitals. While orbitals 4 and 5 are flat orbitals. They mainly contribute to the corresponding bands respectively. The orthorhombic-monoclinic structure phase transition will occur as indicated by black arrows in a low temperature.
   }
   \label{fig1}
\end{figure}

\begin{figure*}
   \centering
    \includegraphics[width=0.85\linewidth]{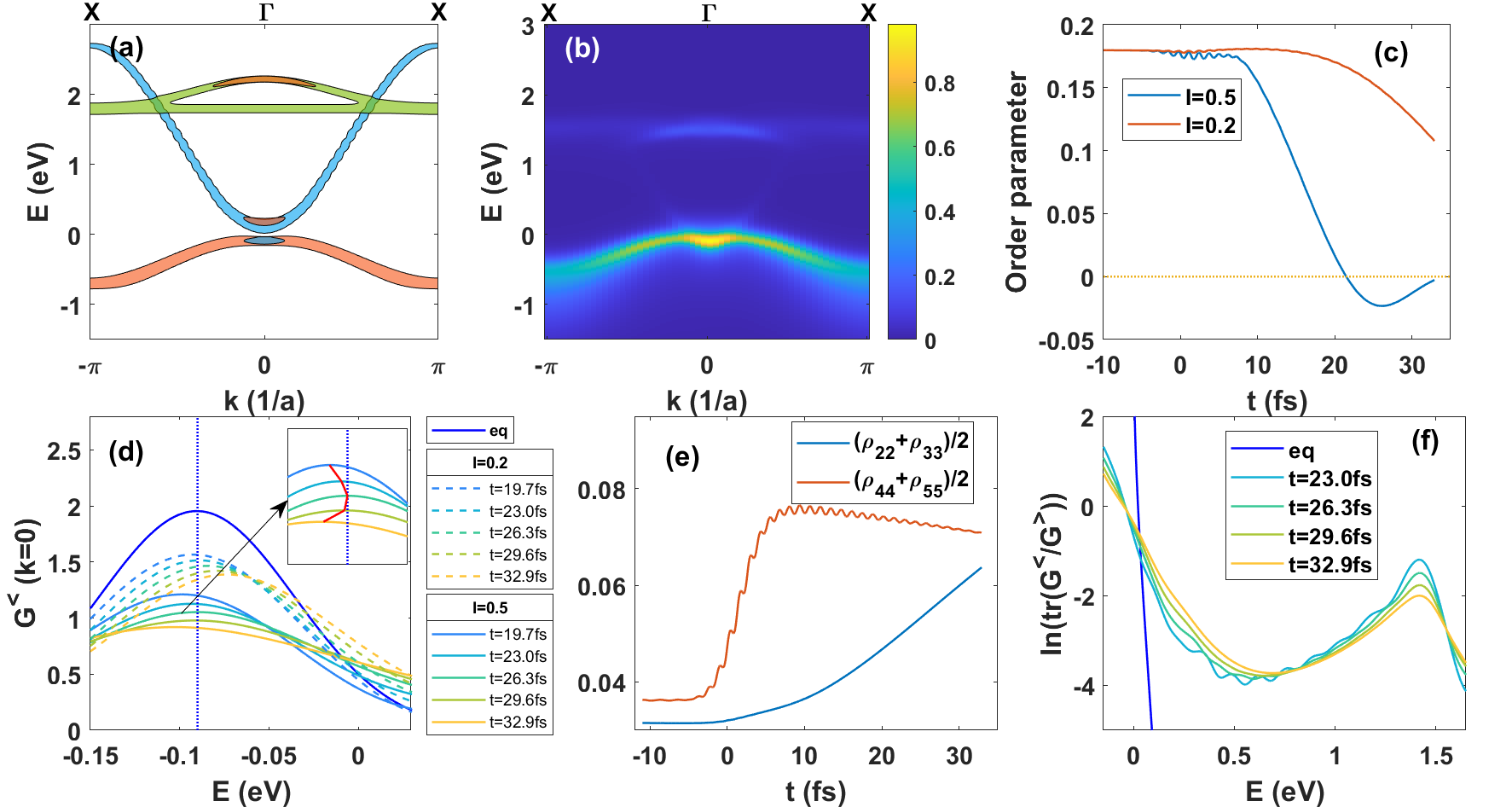}
   \caption{(a) The contribution of different orbitals to the equilibrium spectral function. Colors are the same as those in Fig.~\ref{fig1}. (b) The photoemission spectrum for the case of $I=0.5$ at $t=29.6~{\rm fs}$. (c) The time dependence of the order parameter. (d) The photoemission spectrum at the $\Gamma$ point for two different cases at different times. `eq` represents the result of the equilibrium state. The red line indicates the shift of the peak for the case of $I=0.5$. (e) The time dependence of the average local occupation (diagonal terms of density matrix at $x=0$) of upper bands and flat bands for the case of $I=0.5$. (f) $\ln[{\rm tr}(G^</G^>)]$ at different times for the case of $I=0.5$.
   }
   \label{fig2}
\end{figure*}

In Fig.~\ref{fig1}(a), we display the original energy bands 
obtained from density-functional theory~(DFT) calculations, categorized into three classes: lower bands, upper bands, and flat bands. To construct the minimal model, we select five Wannier orbitals from three atoms within a unit cell, as depicted in Fig.~\ref{fig1}(c). The Hamiltonian without the external field can be expressed as 
\begin{equation}
   H_{\rm eq}=\sum_{k,\alpha,\alpha',\sigma}\epsilon_{k,\alpha,\alpha'}c^{\dagger}_{k,\alpha,\sigma}c_{k,\alpha',\sigma}+H_U+H_V,
 \label{H_eq}
\end{equation}
where the first part is the single-particle Hamiltonian and the others are many-body interactions. Here, we exclusively consider the density-density interactions between orbitals residing on the same atom and on nearest-neighbor $\ce{Ta}$-$\ce{Ni}$ atoms. 
\begin{align}
   H_U=&\sum_{i,\alpha}U_{\alpha}n_{i,\alpha,\uparrow}n_{i,\alpha,\downarrow},\\
   H_V=&\sum_{i,\sigma,\sigma'}\left[V_{\ce{Ta}}(n_{i,2,\sigma}n_{i,4,\sigma'}+n_{i,3,\sigma}n_{i,5,\sigma'}) \nonumber \right.\\
   &+V_{\ce{TaNi}}~n_{i,1,\sigma}(n_{i,2,\sigma'}+n_{i,3,\sigma'}+n_{i,4,\sigma'}+n_{i,5,\sigma'}\nonumber\\
   &\left.+n_{i+1,2,\sigma'}+n_{i+1,4,\sigma'}+n_{i-1,3,\sigma'}+n_{i-1,5,\sigma'})\right],
 \label{H_UV}
\end{align}
where $i$ denotes the cell's position in $\Gamma$\text{-}$X$ direction, $\alpha$ represents the orbital label and $\sigma$ represents the spin. A detailed analysis of both the real material and the model is provided in Supplemental Materials~\cite{SM}. We define the order parameter $\phi$ by the off-diagonal terms of the density matrix $\rho_{\alpha\alpha'}(x)$ in the real space~\cite{mazza2020nature}
\begin{equation}
   \phi=\frac{\left | \rho_{12}(1) \right | + \left | \rho_{13}(-1) \right |-\left | \rho_{12}(0) \right |-\left | \rho_{13}(0) \right |}{4},
 \label{OrderParameter}
\end{equation}
which is close to zero in the orthorhombic phase and has a non-zero value in the monoclinic phase. In Fig.~\ref{fig1}(b), we present the equilibrium spectral function (i.e., imaginary part of the retarded Green's function $G^{\rm ret}$) of this model in the ordered phase at a temperature $T=116~{\rm K}$. We plot the contributions from different orbitals for the equilibrium spectral function in Fig.~\ref{fig2}(a), both the band inversion between upper and lower bands and the strong hybridization between flat and lower bands can be clearly observed near the $\Gamma$ point.

To simulate the laser-induced non-equilibrium dynamics, we use the length gauge to describe the coupling of the light and the matter in the Hamiltonian~\cite{golevz2019multiband,yue2020structure,silva2019high,li2020electromagnetic}. Supposing the position operator is diagonal on the basis of Wannier functions $\left<0\alpha|r|R\alpha'\right>=\delta_{0R}\delta_{\alpha\alpha'}\tau_{\alpha}$, the Hamiltonian in the presence of a laser field can be written as
\begin{equation}
   H(k)=H_{\rm eq}(k-A)+\sum_{\alpha} E\cdot\tau_{\alpha},
 \label{H_neq}
\end{equation}
where $A$ and $E$ represent the vector potential and electric field of the laser pulse respectively, $\tau_{\alpha}$ denotes the center of the orbital $\alpha$ within the unit cell. 

First, we will analyze cases in the time domain. The electric field of the laser pulse we use is given by $E(t)=\sqrt{I}\exp(-t^2/{\tau}^2)\sin(\omega t)$, with specific parameters $\tau=4.5~{\rm fs}$ and $\omega=1.55~{\rm eV}$ and intensity $I$ in arbitrary units. For the case of $I=0.5$, its snapshot of photoemission spectrum (i.e., the lesser Green's function $G^<$) at $t=29.6~{\rm fs}$ is plotted in Fig.~\ref{fig2}(b). It is evident that some electrons are trapped in the flat band, which has been observed in experiments~\cite{mor2022ultrafast}. To show the situation near the Fermi surface more clearly, Fig.~\ref{fig2}(d) shows zoomed-in photoemission spectrum slices at $k=0$ for various time points in the two cases of $I=0.5$ and $I=0.2$. These two figures reveal that the valence band experiences significant depletion and broadening.

Focusing on the peak's position evolution in Fig.~\ref{fig2}(d), we observe that its energy in the case of $I=0.5$ consistently remains below that of the equilibrium state. In the case of $I=0.2$, the peak moves upwards. We will analyze the intensity dependence after and concentrate on the $I=0.5$ case for now. To identify the underlying cause of the downward shift, we plot the population of upper and flat bands against time in Figs.~\ref{fig2}(e). There is a significant transfer from lower bands to flat bands during the laser pulse, which can be attributed to the resonance between the $\Gamma$ point and flat bands and the strong hybridization between the two bands. After the laser pulse, the gradual depletion of flat bands suggests that the trapped state in flat bands is long-lasting, its lifetime can be estimated as 200~{\rm fs} by a linear extrapolation. The relaxation process also induces the transfer of population from the lower band to upper bands. Assuming the population transfer from lower bands to flat bands is $\Delta n_1$ and that from lower bands to upper bands is $\Delta n_2$, one can express the Hartree shifts~\cite{cilento2018dynamics} of the bare lower and upper bands as
\begin{align}
   \Delta E_{\rm upper}=&(\frac{V_{\ce{Ta}}}{2}-2V_{\ce{TaNi}})\Delta n_1+(\frac{U_{\ce{Ta}}}{4}-2V_{\ce{TaNi}})\Delta n_2,\nonumber\\
   \Delta E_{\rm lower}=&(2V_{\ce{TaNi}}-\frac{U_{\ce{Ni}}}{2})\times(\Delta n_1+\Delta n_2).
 \label{HartreeShift}
\end{align}
The coefficients come from the fact that the local interaction exists only between electrons with different spins and one Ni atom is accompanied by two Ta atoms in a unit cell. According to the parameters  from DFT~\cite{SM}, it holds true that $\Delta E_{\rm lower}>0$ and $\Delta E_{\rm upper}<0$ for any positive values $\Delta n_1$ and $\Delta n_2$. As TNS is a topological insulator with band inversions~\cite{ma2022ta}, its valence top is predominantly contributed by upper orbitals, resulting in a negative Hartree shift $\Delta E_{\rm upper}$.  In addition to the contribution from the Hartree shift, the peak position is also influenced by the order parameter dynamics~(the screened Fock term). A photo-induced reduction of the order parameter leads to the collapse of the energy gap~(mark the difference with the BEC case~\cite{golevz2022unveiling}) and an upward shift of the valence band. The band shift near the $\Gamma$ point is sensitive to the competition of the Hartree shift and the collapse effect. In the case of $I=0.5$, the Hartree shift prevails. Similar competition has been discussed in other materials~\cite{spataru2004ab}. The flat band plays an important role in the balance of two contributions. In Fig.~\ref{fig2}(f), the ratio of the lesser  and the greater Green's function $\ln[{\rm tr}(G^</G^>)]$ is plotted for different times and its slope provides an indication of the effective inverse temperature using the fluctuation-dissipation theorem. The symmetry-induced limited hybridization between flat and upper orbitals induces the long-lived states~\cite{SM} in flat bands and slows down the thermalization and the collapse near the Fermi surface, which makes two contributions comparable. We emphasize that our simulations only cover a brief period immediately following the pulse to illustrate the potential for a downward shift. For longer-term evolution, the effect introduced by phonons becomes significant~\cite{hu2022tracking,mor2018}, which is beyond the scope of this study.

\begin{figure}
   \centering
    \includegraphics[width=\linewidth]{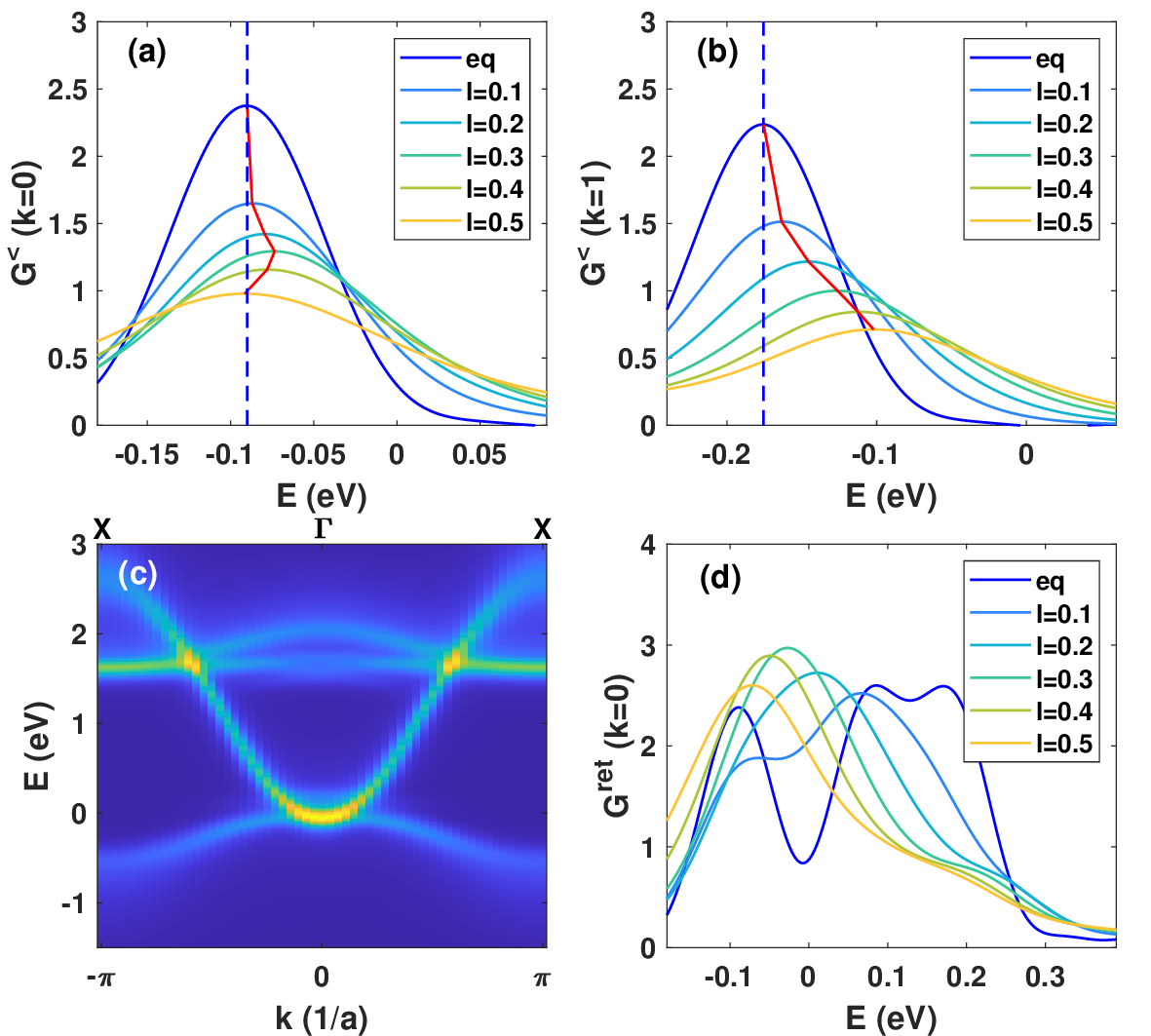}
   \caption{The photoemission spectrum at $k=0$~(a) and $k=1/a$~(b) at $t=29.6~{\rm fs}$ for different laser intensities $I$. (c) The spectral function at $t=29.6~{\rm fs}$ for the laser intensity $I=0.5$. (d) The spectral function at $\Gamma$ and $t=29.6~{\rm fs}$ for different laser intensities.
   }
   \label{fig3}
\end{figure}

To examine the anomalous intensity-dependence and momentum-dependence band shifts, we fix the observed time at $t=29.6~{\rm fs}$ while adjusting the intensity of the pump pulse. The photoemission spectra at $k=0$ and $k=1/a$ for $a$ being the lattice distance are illustrated in Figs.~\ref{fig3} (a) and (b), respectively. A monotonically upward trend in the peak position at $k=1/a$ is observed with the increase of laser intensity. In contrast, at $k=0$, the peak position ascends at a lower laser intensity and declines at a higher laser intensity. This observation is consistent with the experimental findings~\cite{mor2017ultrafast}. The anomalous momentum-dependent band renormalization can be explained by distinct orbital contributions at different momenta. The valence band at $k=1/a$ is primarily contributed by the lower orbital, so the Hartree shift $\Delta E_{\rm lower}$ in Eq.~(\ref{HartreeShift}) and the collapse effect both lift the band. To elucidate the reason for anomalous intensity-dependence at $k=0$, we depict the spectral function for $I=0.5$ in Fig.~\ref{fig3}(c) and provide slices of the spectral function at $k=0$ for varying laser intensities in Fig.~\ref{fig3}(d). With an increase in the intensity of the pump laser, the three peaks observed in the equilibrium spectral function merge, and the energy gap gradually diminishes, illustrating the aforementioned collapse effect. When the intensity is lower than $I=0.3$, the collapse effect is more pronounced than the Hartree shift, causing the band to move upwards. For larger intensities, the collapse effect saturates, and the Hartree shift dominates, causing the band to move downwards. The anomalous intensity-dependent change actually reflects the nonlinear response to the laser field. As the laser intensity increases, the dominated effect may switch due to different nonlinear thresholds. Similar anomalous phenomena have been observed in the absorption peak of other two-dimensional materials with excitonic structures~\cite{jiang2018photo,bera2021atomlike}.

\begin{figure}
   \centering
    \includegraphics[width=\linewidth]{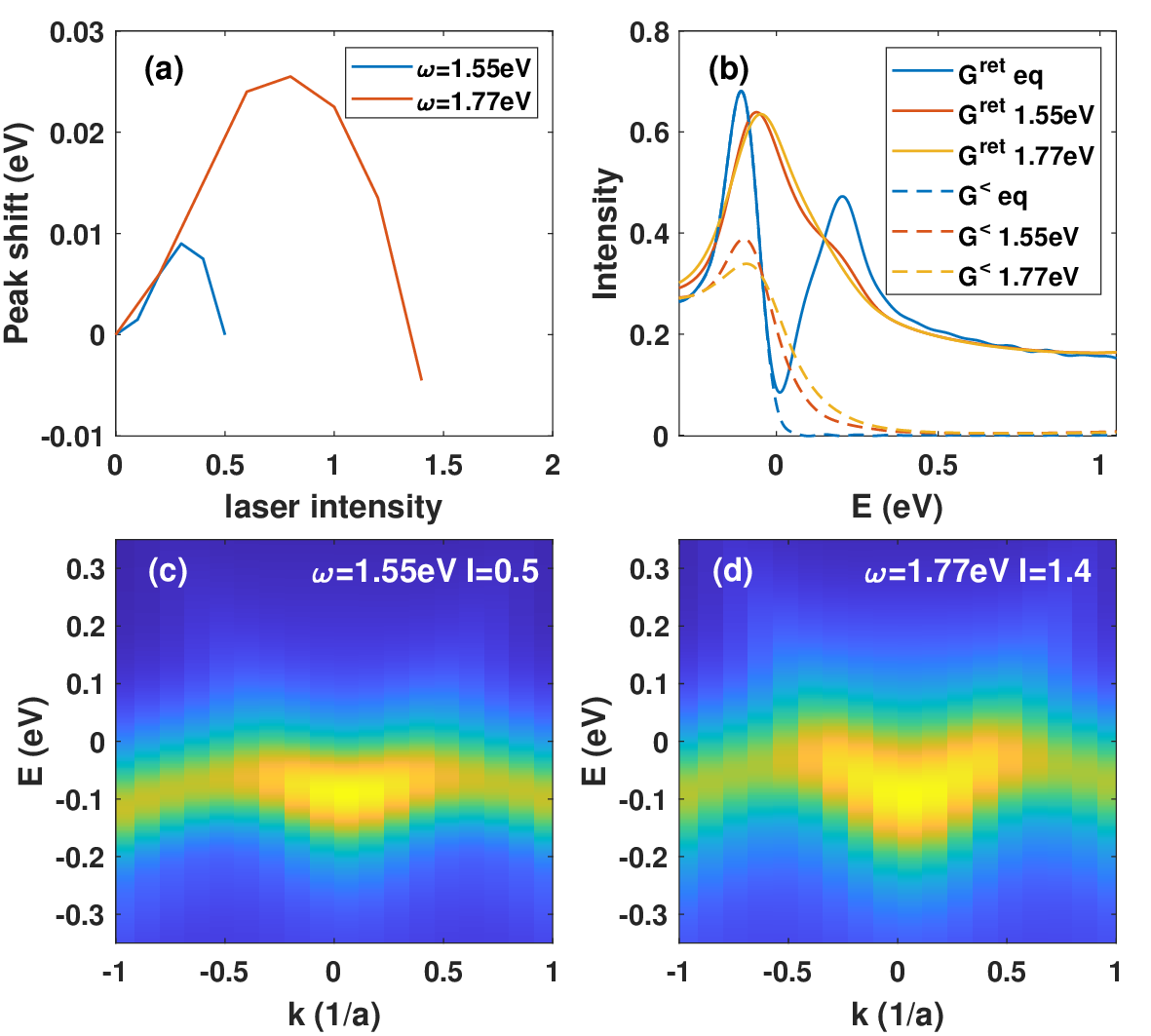}
   \caption{(a) The laser intensity dependence of the shift of the photoemission spectrum peak at $k=0$ and $t=26.3~{\rm fs}$ for two different frequencies. (b) The photoemission spectrum and the spectral function for the equilibrium state, the case of $\omega=1.55~{\rm eV}$, $I=0.5$ and the case of $\omega=1.77~{\rm eV}$, $I=1.4$. (c) The photoemission spectrum near the $\Gamma$ point at $t=26.3~{\rm fs}$ for the case of $I=0.5$, $\omega=1.55~{\rm eV}$. (d) Same as (c) but for the case of $I=1.4$, $\omega=1.77~{\rm eV}$.
   }
   \label{fig4}
\end{figure}

Note that not all the experiments report the downward shift of the valence band top. An example can be found in Ref.~\cite{tang2020non}. We attribute this discrepancy to the use of a pump laser with a frequency $\omega=1.77~{\rm eV}$. To investigate the effect of frequency, we perform numerical calculations for the laser frequency $\omega=1.77~{\rm eV}$ to compare with those results with $\omega=1.55~{\rm eV}$. The peak shift results for both frequencies are plotted against with the laser intensities in Fig.~\ref{fig4}(a). The case with the frequency $\omega=1.55~{\rm eV}$ exhibits a downward shift behavior at significantly lower intensities. To investigate the condition under which the downward shift occurs, we select two examples: $\omega=1.55~{\rm eV}$ with $I=0.5$ and $\omega=1.77~{\rm eV}$ with $I=1.4$. The momentum-integrated spectrum functions and photoemission spectra at $26.3~{\rm fs}$ for both examples are shown in Fig.~\ref{fig4}(b) and the momentum-resolved photoemission spectra in Figs.~\ref{fig4}(c) and (d). The distortion of the valence band for $\omega=1.55~{\rm eV}$ is considerably smaller than that in the high-frequency case, consistent with observations in Refs.~\cite{tang2020non,baldini2023spontaneous}. Although we still observe a downward shift for $\omega=1.77~{\rm eV}$ at a higher intensity, the strongly distorted electronic structure under this condition raises concerns about the validity of our model in that condition. The selective excitation exemplified in this paragraph is common in other photo-doped materials~\cite{pagliara2011photoinduced,mitrano2016possible,ligges2018ultrafast} and  usually indicates the complex excited state structure, such as the Van Hove singularity. 

Although we have focused on TNS, this work introduces new and generic strategies for the bandgap renormalization in diverse materials. Here, we gather necessary properties of materials for the anomalous band renormalization. Firstly, a strong Coulomb interaction proves to be pivotal for observing the photo-induced band renormalization. Consequently, low-dimensional materials stands out as promising platforms due to their less screened interaction. Additionally, the existence of long-lived non-equilibrium states is crucial and in this work we use selective excitation of flatbands to create long-lived states. For the anomalous pump parameter dependence, the existence of competing mechanisms is a key point. Notably, the band inversion may induce anomalies in momentum dependence, suggesting that topological insulators with band inversions may exhibit similar anomalies in the momentum space. The intensity dependence implies the nonlinear responses, while frequency dependence implies a specific excited state structure. Past experiments highlight materials with excitonic structures as candidates likely to display anomalous dependence on the laser parameter in photo-induced phenomena~\cite{jiang2018photo,bera2021atomlike}. 

In conclusion, we have developed a concise model for TNS and conducted systematic simulations to explore its laser-induced non-equilibrium electronic dynamics. The anomalous parameter dependence of the photoemission spectra are carefully analyzed. This investigation not only sheds light on the anomalous band renormalization in TNS, but also provides insights for identifying other materials with similar features and designing laser protocols to unveil anomalous photo-induced phenomena. As low-dimensional materials and systems with Moire patterns~\cite{Li2021,Tang2020,Cao2018,Cao2018a} generically poses flat bands we envision these systems as
natural candidates for long-lived photo-modification of the bandstructure.



\begin{acknowledgements}
   L. G. is grateful to G. Mazza, M. Rösner, Z. Sun and H. Yang for fruitful discussions. L. G., X. L., J. Z. and L. -Y. P. acknowledge support from No. 12234002 and No. 92250303 of the National Natural Science Foundation of China. D. G. acknowledge support from No. P1-0044, No. J1-2455 and No. MN-0016-106 of the Slovenian Research Agency (ARRS).
\end{acknowledgements}

\appendix


\begin{thebibliography}{69}%
   \makeatletter
   \providecommand \@ifxundefined [1]{%
    \@ifx{#1\undefined}
   }%
   \providecommand \@ifnum [1]{%
    \ifnum #1\expandafter \@firstoftwo
    \else \expandafter \@secondoftwo
    \fi
   }%
   \providecommand \@ifx [1]{%
    \ifx #1\expandafter \@firstoftwo
    \else \expandafter \@secondoftwo
    \fi
   }%
   \providecommand \natexlab [1]{#1}%
   \providecommand \enquote  [1]{``#1''}%
   \providecommand \bibnamefont  [1]{#1}%
   \providecommand \bibfnamefont [1]{#1}%
   \providecommand \citenamefont [1]{#1}%
   \providecommand \href@noop [0]{\@secondoftwo}%
   \providecommand \href [0]{\begingroup \@sanitize@url \@href}%
   \providecommand \@href[1]{\@@startlink{#1}\@@href}%
   \providecommand \@@href[1]{\endgroup#1\@@endlink}%
   \providecommand \@sanitize@url [0]{\catcode `\\12\catcode `\$12\catcode
     `\&12\catcode `\#12\catcode `\^12\catcode `\_12\catcode `\%12\relax}%
   \providecommand \@@startlink[1]{}%
   \providecommand \@@endlink[0]{}%
   \providecommand \url  [0]{\begingroup\@sanitize@url \@url }%
   \providecommand \@url [1]{\endgroup\@href {#1}{\urlprefix }}%
   \providecommand \urlprefix  [0]{URL }%
   \providecommand \Eprint [0]{\href }%
   \providecommand \doibase [0]{http://dx.doi.org/}%
   \providecommand \selectlanguage [0]{\@gobble}%
   \providecommand \bibinfo  [0]{\@secondoftwo}%
   \providecommand \bibfield  [0]{\@secondoftwo}%
   \providecommand \translation [1]{[#1]}%
   \providecommand \BibitemOpen [0]{}%
   \providecommand \bibitemStop [0]{}%
   \providecommand \bibitemNoStop [0]{.\EOS\space}%
   \providecommand \EOS [0]{\spacefactor3000\relax}%
   \providecommand \BibitemShut  [1]{\csname bibitem#1\endcsname}%
   \let\auto@bib@innerbib\@empty
   \bibitem [{\citenamefont {Oka}\ and\ \citenamefont
     {Kitamura}(2019)}]{oka2019floquet}%
     \BibitemOpen
     \bibfield  {author} {\bibinfo {author} {\bibfnamefont {T.}~\bibnamefont
     {Oka}}\ and\ \bibinfo {author} {\bibfnamefont {S.}~\bibnamefont {Kitamura}},\
     }\href@noop {} {\bibfield  {journal} {\bibinfo  {journal} {Annu. Rev.
     Condens. Matter Phys.}\ }\textbf {\bibinfo {volume} {10}},\ \bibinfo {pages}
     {387} (\bibinfo {year} {2019})}\BibitemShut {NoStop}%
   \bibitem [{\citenamefont {De~La~Torre}\ \emph {et~al.}(2021)\citenamefont
     {De~La~Torre}, \citenamefont {Kennes}, \citenamefont {Claassen},
     \citenamefont {Gerber}, \citenamefont {McIver},\ and\ \citenamefont
     {Sentef}}]{de2021colloquium}%
     \BibitemOpen
     \bibfield  {author} {\bibinfo {author} {\bibfnamefont {A.}~\bibnamefont
     {De~La~Torre}}, \bibinfo {author} {\bibfnamefont {D.~M.}\ \bibnamefont
     {Kennes}}, \bibinfo {author} {\bibfnamefont {M.}~\bibnamefont {Claassen}},
     \bibinfo {author} {\bibfnamefont {S.}~\bibnamefont {Gerber}}, \bibinfo
     {author} {\bibfnamefont {J.~W.}\ \bibnamefont {McIver}}, \ and\ \bibinfo
     {author} {\bibfnamefont {M.~A.}\ \bibnamefont {Sentef}},\ }\href@noop {}
     {\bibfield  {journal} {\bibinfo  {journal} {Rev. Mod. Phys.}\ }\textbf
     {\bibinfo {volume} {93}},\ \bibinfo {pages} {041002} (\bibinfo {year}
     {2021})}\BibitemShut {NoStop}%
   \bibitem [{\citenamefont {Aoki}\ \emph {et~al.}(2014)\citenamefont {Aoki},
     \citenamefont {Tsuji}, \citenamefont {Eckstein}, \citenamefont {Kollar},
     \citenamefont {Oka},\ and\ \citenamefont {Werner}}]{aoki2014nonequilibrium}%
     \BibitemOpen
     \bibfield  {author} {\bibinfo {author} {\bibfnamefont {H.}~\bibnamefont
     {Aoki}}, \bibinfo {author} {\bibfnamefont {N.}~\bibnamefont {Tsuji}},
     \bibinfo {author} {\bibfnamefont {M.}~\bibnamefont {Eckstein}}, \bibinfo
     {author} {\bibfnamefont {M.}~\bibnamefont {Kollar}}, \bibinfo {author}
     {\bibfnamefont {T.}~\bibnamefont {Oka}}, \ and\ \bibinfo {author}
     {\bibfnamefont {P.}~\bibnamefont {Werner}},\ }\href@noop {} {\bibfield
     {journal} {\bibinfo  {journal} {Rev. Mod. Phys.}\ }\textbf {\bibinfo {volume}
     {86}},\ \bibinfo {pages} {779} (\bibinfo {year} {2014})}\BibitemShut
     {NoStop}%
   \bibitem [{\citenamefont {Murakami}\ \emph {et~al.}(2023)\citenamefont
     {Murakami}, \citenamefont {Gole{\v{z}}}, \citenamefont {Eckstein},\ and\
     \citenamefont {Werner}}]{murakami2023photo}%
     \BibitemOpen
     \bibfield  {author} {\bibinfo {author} {\bibfnamefont {Y.}~\bibnamefont
     {Murakami}}, \bibinfo {author} {\bibfnamefont {D.}~\bibnamefont
     {Gole{\v{z}}}}, \bibinfo {author} {\bibfnamefont {M.}~\bibnamefont
     {Eckstein}}, \ and\ \bibinfo {author} {\bibfnamefont {P.}~\bibnamefont
     {Werner}},\ }\href@noop {} {\bibfield  {journal} {\bibinfo  {journal}
     {arXiv:2310.05201}\ } (\bibinfo {year} {2023})}\BibitemShut {NoStop}%
   \bibitem [{\citenamefont {Disa}\ \emph {et~al.}(2023)\citenamefont {Disa},
     \citenamefont {Curtis}, \citenamefont {Fechner}, \citenamefont {Liu},
     \citenamefont {von Hoegen}, \citenamefont {F{\"o}rst}, \citenamefont {Nova},
     \citenamefont {Narang}, \citenamefont {Maljuk}, \citenamefont {Boris} \emph
     {et~al.}}]{disa2023photo}%
     \BibitemOpen
     \bibfield  {author} {\bibinfo {author} {\bibfnamefont {A.}~\bibnamefont
     {Disa}}, \bibinfo {author} {\bibfnamefont {J.}~\bibnamefont {Curtis}},
     \bibinfo {author} {\bibfnamefont {M.}~\bibnamefont {Fechner}}, \bibinfo
     {author} {\bibfnamefont {A.}~\bibnamefont {Liu}}, \bibinfo {author}
     {\bibfnamefont {A.}~\bibnamefont {von Hoegen}}, \bibinfo {author}
     {\bibfnamefont {M.}~\bibnamefont {F{\"o}rst}}, \bibinfo {author}
     {\bibfnamefont {T.}~\bibnamefont {Nova}}, \bibinfo {author} {\bibfnamefont
     {P.}~\bibnamefont {Narang}}, \bibinfo {author} {\bibfnamefont
     {A.}~\bibnamefont {Maljuk}}, \bibinfo {author} {\bibfnamefont
     {A.}~\bibnamefont {Boris}},  \emph {et~al.},\ }\href@noop {} {\bibfield
     {journal} {\bibinfo  {journal} {Nature}\ }\textbf {\bibinfo {volume} {617}},\
     \bibinfo {pages} {73} (\bibinfo {year} {2023})}\BibitemShut {NoStop}%
   \bibitem [{\citenamefont {Mankowsky}\ \emph {et~al.}(2014)\citenamefont
     {Mankowsky}, \citenamefont {Subedi}, \citenamefont {F{\"o}rst}, \citenamefont
     {Mariager}, \citenamefont {Chollet}, \citenamefont {Lemke}, \citenamefont
     {Robinson}, \citenamefont {Glownia}, \citenamefont {Minitti}, \citenamefont
     {Frano} \emph {et~al.}}]{mankowsky2014nonlinear}%
     \BibitemOpen
     \bibfield  {author} {\bibinfo {author} {\bibfnamefont {R.}~\bibnamefont
     {Mankowsky}}, \bibinfo {author} {\bibfnamefont {A.}~\bibnamefont {Subedi}},
     \bibinfo {author} {\bibfnamefont {M.}~\bibnamefont {F{\"o}rst}}, \bibinfo
     {author} {\bibfnamefont {S.~O.}\ \bibnamefont {Mariager}}, \bibinfo {author}
     {\bibfnamefont {M.}~\bibnamefont {Chollet}}, \bibinfo {author} {\bibfnamefont
     {H.}~\bibnamefont {Lemke}}, \bibinfo {author} {\bibfnamefont {J.~S.}\
     \bibnamefont {Robinson}}, \bibinfo {author} {\bibfnamefont {J.~M.}\
     \bibnamefont {Glownia}}, \bibinfo {author} {\bibfnamefont {M.~P.}\
     \bibnamefont {Minitti}}, \bibinfo {author} {\bibfnamefont {A.}~\bibnamefont
     {Frano}},  \emph {et~al.},\ }\href@noop {} {\bibfield  {journal} {\bibinfo
     {journal} {Nature}\ }\textbf {\bibinfo {volume} {516}},\ \bibinfo {pages}
     {71} (\bibinfo {year} {2014})}\BibitemShut {NoStop}%
   \bibitem [{\citenamefont {Mitrano}\ \emph {et~al.}(2016)\citenamefont
     {Mitrano}, \citenamefont {Cantaluppi}, \citenamefont {Nicoletti},
     \citenamefont {Kaiser}, \citenamefont {Perucchi}, \citenamefont {Lupi},
     \citenamefont {Di~Pietro}, \citenamefont {Pontiroli}, \citenamefont
     {Ricc{\`o}}, \citenamefont {Clark} \emph {et~al.}}]{mitrano2016possible}%
     \BibitemOpen
     \bibfield  {author} {\bibinfo {author} {\bibfnamefont {M.}~\bibnamefont
     {Mitrano}}, \bibinfo {author} {\bibfnamefont {A.}~\bibnamefont {Cantaluppi}},
     \bibinfo {author} {\bibfnamefont {D.}~\bibnamefont {Nicoletti}}, \bibinfo
     {author} {\bibfnamefont {S.}~\bibnamefont {Kaiser}}, \bibinfo {author}
     {\bibfnamefont {A.}~\bibnamefont {Perucchi}}, \bibinfo {author}
     {\bibfnamefont {S.}~\bibnamefont {Lupi}}, \bibinfo {author} {\bibfnamefont
     {P.}~\bibnamefont {Di~Pietro}}, \bibinfo {author} {\bibfnamefont
     {D.}~\bibnamefont {Pontiroli}}, \bibinfo {author} {\bibfnamefont
     {M.}~\bibnamefont {Ricc{\`o}}}, \bibinfo {author} {\bibfnamefont {S.~R.}\
     \bibnamefont {Clark}},  \emph {et~al.},\ }\href@noop {} {\bibfield  {journal}
     {\bibinfo  {journal} {Nature}\ }\textbf {\bibinfo {volume} {530}},\ \bibinfo
     {pages} {461} (\bibinfo {year} {2016})}\BibitemShut {NoStop}%
   \bibitem [{\citenamefont {Fausti}\ \emph {et~al.}(2011)\citenamefont {Fausti},
     \citenamefont {Tobey}, \citenamefont {Dean}, \citenamefont {Kaiser},
     \citenamefont {Dienst}, \citenamefont {Hoffmann}, \citenamefont {Pyon},
     \citenamefont {Takayama}, \citenamefont {Takagi},\ and\ \citenamefont
     {Cavalleri}}]{Fausti11}%
     \BibitemOpen
     \bibfield  {author} {\bibinfo {author} {\bibfnamefont {D.}~\bibnamefont
     {Fausti}}, \bibinfo {author} {\bibfnamefont {R.~I.}\ \bibnamefont {Tobey}},
     \bibinfo {author} {\bibfnamefont {N.}~\bibnamefont {Dean}}, \bibinfo {author}
     {\bibfnamefont {S.}~\bibnamefont {Kaiser}}, \bibinfo {author} {\bibfnamefont
     {A.}~\bibnamefont {Dienst}}, \bibinfo {author} {\bibfnamefont {M.~C.}\
     \bibnamefont {Hoffmann}}, \bibinfo {author} {\bibfnamefont {S.}~\bibnamefont
     {Pyon}}, \bibinfo {author} {\bibfnamefont {T.}~\bibnamefont {Takayama}},
     \bibinfo {author} {\bibfnamefont {H.}~\bibnamefont {Takagi}}, \ and\ \bibinfo
     {author} {\bibfnamefont {A.}~\bibnamefont {Cavalleri}},\ }\href {\doibase
     10.1126/science.1197294} {\bibfield  {journal} {\bibinfo  {journal}
     {Science}\ }\textbf {\bibinfo {volume} {331}},\ \bibinfo {pages} {189}
     (\bibinfo {year} {2011})},\ \Eprint
     {http://arxiv.org/abs/https://www.science.org/doi/pdf/10.1126/science.1197294}
     {https://www.science.org/doi/pdf/10.1126/science.1197294} \BibitemShut
     {NoStop}%
   \bibitem [{\citenamefont {Stojchevska}\ \emph {et~al.}(2014)\citenamefont
     {Stojchevska}, \citenamefont {Vaskivskyi}, \citenamefont {Mertelj},
     \citenamefont {Kusar}, \citenamefont {Svetin}, \citenamefont {Brazovskii},\
     and\ \citenamefont {Mihailovic}}]{Stojchevska2014}%
     \BibitemOpen
     \bibfield  {author} {\bibinfo {author} {\bibfnamefont {L.}~\bibnamefont
     {Stojchevska}}, \bibinfo {author} {\bibfnamefont {I.}~\bibnamefont
     {Vaskivskyi}}, \bibinfo {author} {\bibfnamefont {T.}~\bibnamefont {Mertelj}},
     \bibinfo {author} {\bibfnamefont {P.}~\bibnamefont {Kusar}}, \bibinfo
     {author} {\bibfnamefont {D.}~\bibnamefont {Svetin}}, \bibinfo {author}
     {\bibfnamefont {A.}~\bibnamefont {Brazovskii}}, \ and\ \bibinfo {author}
     {\bibfnamefont {D.}~\bibnamefont {Mihailovic}},\ }\href {\doibase
     10.1126/science.1241591} {\bibfield  {journal} {\bibinfo  {journal}
     {Science}\ ,\ \bibinfo {pages} {177}} (\bibinfo {year} {2014})}\BibitemShut
     {NoStop}%
   \bibitem [{\citenamefont {Vaskivskyi}\ \emph {et~al.}(2016)\citenamefont
     {Vaskivskyi}, \citenamefont {Mihailovic}, \citenamefont {Brazovskii},
     \citenamefont {Gospodaric}, \citenamefont {Mertelj}, \citenamefont {Svetin},
     \citenamefont {Sutar},\ and\ \citenamefont {Mihailovic}}]{Vaskivskyi2016}%
     \BibitemOpen
     \bibfield  {author} {\bibinfo {author} {\bibfnamefont {I.}~\bibnamefont
     {Vaskivskyi}}, \bibinfo {author} {\bibfnamefont {I.~A.}\ \bibnamefont
     {Mihailovic}}, \bibinfo {author} {\bibfnamefont {S.}~\bibnamefont
     {Brazovskii}}, \bibinfo {author} {\bibfnamefont {J.}~\bibnamefont
     {Gospodaric}}, \bibinfo {author} {\bibfnamefont {T.}~\bibnamefont {Mertelj}},
     \bibinfo {author} {\bibfnamefont {D.}~\bibnamefont {Svetin}}, \bibinfo
     {author} {\bibfnamefont {P.}~\bibnamefont {Sutar}}, \ and\ \bibinfo {author}
     {\bibfnamefont {D.}~\bibnamefont {Mihailovic}},\ }\href {\doibase
     10.1038/NCOMMS11442} {\bibfield  {journal} {\bibinfo  {journal} {Nature
     Communications}\ ,\ \bibinfo {pages} {11442}} (\bibinfo {year}
     {2016})}\BibitemShut {NoStop}%
   \bibitem [{\citenamefont {Kogar}\ \emph {et~al.}(2019)\citenamefont {Kogar},
     \citenamefont {Zong}, \citenamefont {Dolgirev}, \citenamefont {Shen},
     \citenamefont {Straquadine}, \citenamefont {Bie}, \citenamefont {Wang},
     \citenamefont {Rohwer}, \citenamefont {Tung}, \citenamefont {Yang},
     \citenamefont {Li}, \citenamefont {Yang}, \citenamefont {Weathersby},
     \citenamefont {Park}, \citenamefont {Kozina}, \citenamefont {Sie},
     \citenamefont {Wen}, \citenamefont {Jarillo-Herrero}, \citenamefont {Fisher},
     \citenamefont {Wang},\ and\ \citenamefont {Gedik}}]{kogar2020}%
     \BibitemOpen
     \bibfield  {author} {\bibinfo {author} {\bibfnamefont {A.}~\bibnamefont
     {Kogar}}, \bibinfo {author} {\bibfnamefont {A.}~\bibnamefont {Zong}},
     \bibinfo {author} {\bibfnamefont {P.~E.}\ \bibnamefont {Dolgirev}}, \bibinfo
     {author} {\bibfnamefont {X.}~\bibnamefont {Shen}}, \bibinfo {author}
     {\bibfnamefont {J.}~\bibnamefont {Straquadine}}, \bibinfo {author}
     {\bibfnamefont {Y.-Q.}\ \bibnamefont {Bie}}, \bibinfo {author} {\bibfnamefont
     {X.}~\bibnamefont {Wang}}, \bibinfo {author} {\bibfnamefont {T.}~\bibnamefont
     {Rohwer}}, \bibinfo {author} {\bibfnamefont {I.-C.}\ \bibnamefont {Tung}},
     \bibinfo {author} {\bibfnamefont {Y.}~\bibnamefont {Yang}}, \bibinfo {author}
     {\bibfnamefont {R.}~\bibnamefont {Li}}, \bibinfo {author} {\bibfnamefont
     {J.}~\bibnamefont {Yang}}, \bibinfo {author} {\bibfnamefont {S.}~\bibnamefont
     {Weathersby}}, \bibinfo {author} {\bibfnamefont {S.}~\bibnamefont {Park}},
     \bibinfo {author} {\bibfnamefont {M.~E.}\ \bibnamefont {Kozina}}, \bibinfo
     {author} {\bibfnamefont {E.~J.}\ \bibnamefont {Sie}}, \bibinfo {author}
     {\bibfnamefont {H.}~\bibnamefont {Wen}}, \bibinfo {author} {\bibfnamefont
     {P.}~\bibnamefont {Jarillo-Herrero}}, \bibinfo {author} {\bibfnamefont
     {I.~R.}\ \bibnamefont {Fisher}}, \bibinfo {author} {\bibfnamefont
     {X.}~\bibnamefont {Wang}}, \ and\ \bibinfo {author} {\bibfnamefont
     {N.}~\bibnamefont {Gedik}},\ }\href {\doibase 10.1038/s41567-019-0705-3}
     {\bibfield  {journal} {\bibinfo  {journal} {Nature Physics}\ }\textbf
     {\bibinfo {volume} {16}},\ \bibinfo {pages} {159} (\bibinfo {year}
     {2019})}\BibitemShut {NoStop}%
   \bibitem [{\citenamefont {Zong}\ \emph {et~al.}(2019)\citenamefont {Zong},
     \citenamefont {Dolgirev}, \citenamefont {Kogar}, \citenamefont
     {Erge\ifmmode~\mbox{\c{c}}\else \c{c}\fi{}en}, \citenamefont {Yilmaz},
     \citenamefont {Bie}, \citenamefont {Rohwer}, \citenamefont {Tung},
     \citenamefont {Straquadine}, \citenamefont {Wang}, \citenamefont {Yang},
     \citenamefont {Shen}, \citenamefont {Li}, \citenamefont {Yang}, \citenamefont
     {Park}, \citenamefont {Hoffmann}, \citenamefont {Ofori-Okai}, \citenamefont
     {Kozina}, \citenamefont {Wen}, \citenamefont {Wang}, \citenamefont {Fisher},
     \citenamefont {Jarillo-Herrero},\ and\ \citenamefont {Gedik}}]{Zong2019a}%
     \BibitemOpen
     \bibfield  {author} {\bibinfo {author} {\bibfnamefont {A.}~\bibnamefont
     {Zong}}, \bibinfo {author} {\bibfnamefont {P.~E.}\ \bibnamefont {Dolgirev}},
     \bibinfo {author} {\bibfnamefont {A.}~\bibnamefont {Kogar}}, \bibinfo
     {author} {\bibfnamefont {E.}~\bibnamefont {Erge\ifmmode~\mbox{\c{c}}\else
     \c{c}\fi{}en}}, \bibinfo {author} {\bibfnamefont {M.~B.}\ \bibnamefont
     {Yilmaz}}, \bibinfo {author} {\bibfnamefont {Y.-Q.}\ \bibnamefont {Bie}},
     \bibinfo {author} {\bibfnamefont {T.}~\bibnamefont {Rohwer}}, \bibinfo
     {author} {\bibfnamefont {I.-C.}\ \bibnamefont {Tung}}, \bibinfo {author}
     {\bibfnamefont {J.}~\bibnamefont {Straquadine}}, \bibinfo {author}
     {\bibfnamefont {X.}~\bibnamefont {Wang}}, \bibinfo {author} {\bibfnamefont
     {Y.}~\bibnamefont {Yang}}, \bibinfo {author} {\bibfnamefont {X.}~\bibnamefont
     {Shen}}, \bibinfo {author} {\bibfnamefont {R.}~\bibnamefont {Li}}, \bibinfo
     {author} {\bibfnamefont {J.}~\bibnamefont {Yang}}, \bibinfo {author}
     {\bibfnamefont {S.}~\bibnamefont {Park}}, \bibinfo {author} {\bibfnamefont
     {M.~C.}\ \bibnamefont {Hoffmann}}, \bibinfo {author} {\bibfnamefont {B.~K.}\
     \bibnamefont {Ofori-Okai}}, \bibinfo {author} {\bibfnamefont {M.~E.}\
     \bibnamefont {Kozina}}, \bibinfo {author} {\bibfnamefont {H.}~\bibnamefont
     {Wen}}, \bibinfo {author} {\bibfnamefont {X.}~\bibnamefont {Wang}}, \bibinfo
     {author} {\bibfnamefont {I.~R.}\ \bibnamefont {Fisher}}, \bibinfo {author}
     {\bibfnamefont {P.}~\bibnamefont {Jarillo-Herrero}}, \ and\ \bibinfo {author}
     {\bibfnamefont {N.}~\bibnamefont {Gedik}},\ }\href {\doibase
     10.1103/PhysRevLett.123.097601} {\bibfield  {journal} {\bibinfo  {journal}
     {Phys. Rev. Lett.}\ }\textbf {\bibinfo {volume} {123}},\ \bibinfo {pages}
     {097601} (\bibinfo {year} {2019})}\BibitemShut {NoStop}%
   \bibitem [{\citenamefont {Boschini}\ \emph {et~al.}(2023)\citenamefont
     {Boschini}, \citenamefont {Zonno},\ and\ \citenamefont
     {Damascelli}}]{boschini2023time}%
     \BibitemOpen
     \bibfield  {author} {\bibinfo {author} {\bibfnamefont {F.}~\bibnamefont
     {Boschini}}, \bibinfo {author} {\bibfnamefont {M.}~\bibnamefont {Zonno}}, \
     and\ \bibinfo {author} {\bibfnamefont {A.}~\bibnamefont {Damascelli}},\
     }\href@noop {} {\enquote {\bibinfo {title} {Time- and angle-resolved
     photoemission studies of quantum materials},}\ } (\bibinfo {year} {2023}),\
     \Eprint {http://arxiv.org/abs/2309.03935} {arXiv:2309.03935
     [cond-mat.str-el]} \BibitemShut {NoStop}%
   \bibitem [{\citenamefont {Perfetti}\ \emph {et~al.}(2006)\citenamefont
     {Perfetti}, \citenamefont {Loukakos}, \citenamefont {Lisowski}, \citenamefont
     {Bovensiepen}, \citenamefont {Berger}, \citenamefont {Biermann},
     \citenamefont {Cornaglia}, \citenamefont {Georges},\ and\ \citenamefont
     {Wolf}}]{Perfetti2006}%
     \BibitemOpen
     \bibfield  {author} {\bibinfo {author} {\bibfnamefont {L.}~\bibnamefont
     {Perfetti}}, \bibinfo {author} {\bibfnamefont {P.~A.}\ \bibnamefont
     {Loukakos}}, \bibinfo {author} {\bibfnamefont {M.}~\bibnamefont {Lisowski}},
     \bibinfo {author} {\bibfnamefont {U.}~\bibnamefont {Bovensiepen}}, \bibinfo
     {author} {\bibfnamefont {H.}~\bibnamefont {Berger}}, \bibinfo {author}
     {\bibfnamefont {S.}~\bibnamefont {Biermann}}, \bibinfo {author}
     {\bibfnamefont {P.~S.}\ \bibnamefont {Cornaglia}}, \bibinfo {author}
     {\bibfnamefont {A.}~\bibnamefont {Georges}}, \ and\ \bibinfo {author}
     {\bibfnamefont {M.}~\bibnamefont {Wolf}},\ }\href@noop {} {\bibfield
     {journal} {\bibinfo  {journal} {Phys. Rev. Lett.}\ }\textbf {\bibinfo
     {volume} {97}},\ \bibinfo {pages} {067402} (\bibinfo {year}
     {2006})}\BibitemShut {NoStop}%
   \bibitem [{\citenamefont {Perfetti}\ \emph {et~al.}(2008)\citenamefont
     {Perfetti}, \citenamefont {Loukakos}, \citenamefont {Lisowski}, \citenamefont
     {Bovensiepen}, \citenamefont {Wolf}, \citenamefont {Berger}, \citenamefont
     {Biermann},\ and\ \citenamefont {Georges}}]{Perfetti2008}%
     \BibitemOpen
     \bibfield  {author} {\bibinfo {author} {\bibfnamefont {L.}~\bibnamefont
     {Perfetti}}, \bibinfo {author} {\bibfnamefont {P.~A.}\ \bibnamefont
     {Loukakos}}, \bibinfo {author} {\bibfnamefont {M.}~\bibnamefont {Lisowski}},
     \bibinfo {author} {\bibfnamefont {U.}~\bibnamefont {Bovensiepen}}, \bibinfo
     {author} {\bibfnamefont {M.}~\bibnamefont {Wolf}}, \bibinfo {author}
     {\bibfnamefont {H.}~\bibnamefont {Berger}}, \bibinfo {author} {\bibfnamefont
     {S.}~\bibnamefont {Biermann}}, \ and\ \bibinfo {author} {\bibfnamefont
     {A.}~\bibnamefont {Georges}},\ }\href@noop {} {\bibfield  {journal} {\bibinfo
      {journal} {New J. Phys.}\ }\textbf {\bibinfo {volume} {10}},\ \bibinfo
     {pages} {053019} (\bibinfo {year} {2008})}\BibitemShut {NoStop}%
   \bibitem [{\citenamefont {Ligges}\ \emph
     {et~al.}(2018{\natexlab{a}})\citenamefont {Ligges}, \citenamefont {Avigo},
     \citenamefont {Gole\ifmmode~\check{z}\else \v{z}\fi{}}, \citenamefont
     {Strand}, \citenamefont {Beyazit}, \citenamefont {Hanff}, \citenamefont
     {Diekmann}, \citenamefont {Stojchevska}, \citenamefont {Kall\"ane},
     \citenamefont {Zhou}, \citenamefont {Rossnagel}, \citenamefont {Eckstein},
     \citenamefont {Werner},\ and\ \citenamefont {Bovensiepen}}]{Ligges2018}%
     \BibitemOpen
     \bibfield  {author} {\bibinfo {author} {\bibfnamefont {M.}~\bibnamefont
     {Ligges}}, \bibinfo {author} {\bibfnamefont {I.}~\bibnamefont {Avigo}},
     \bibinfo {author} {\bibfnamefont {D.}~\bibnamefont
     {Gole\ifmmode~\check{z}\else \v{z}\fi{}}}, \bibinfo {author} {\bibfnamefont
     {H.~U.~R.}\ \bibnamefont {Strand}}, \bibinfo {author} {\bibfnamefont
     {Y.}~\bibnamefont {Beyazit}}, \bibinfo {author} {\bibfnamefont
     {K.}~\bibnamefont {Hanff}}, \bibinfo {author} {\bibfnamefont
     {F.}~\bibnamefont {Diekmann}}, \bibinfo {author} {\bibfnamefont
     {L.}~\bibnamefont {Stojchevska}}, \bibinfo {author} {\bibfnamefont
     {M.}~\bibnamefont {Kall\"ane}}, \bibinfo {author} {\bibfnamefont
     {P.}~\bibnamefont {Zhou}}, \bibinfo {author} {\bibfnamefont {K.}~\bibnamefont
     {Rossnagel}}, \bibinfo {author} {\bibfnamefont {M.}~\bibnamefont {Eckstein}},
     \bibinfo {author} {\bibfnamefont {P.}~\bibnamefont {Werner}}, \ and\ \bibinfo
     {author} {\bibfnamefont {U.}~\bibnamefont {Bovensiepen}},\ }\href {\doibase
     10.1103/PhysRevLett.120.166401} {\bibfield  {journal} {\bibinfo  {journal}
     {Phys. Rev. Lett.}\ }\textbf {\bibinfo {volume} {120}},\ \bibinfo {pages}
     {166401} (\bibinfo {year} {2018}{\natexlab{a}})}\BibitemShut {NoStop}%
   \bibitem [{\citenamefont {Mor}\ \emph {et~al.}(2017)\citenamefont {Mor},
     \citenamefont {Herzog}, \citenamefont {Gole{\v{z}}}, \citenamefont {Werner},
     \citenamefont {Eckstein}, \citenamefont {Katayama}, \citenamefont {Nohara},
     \citenamefont {Takagi}, \citenamefont {Mizokawa}, \citenamefont {Monney}
     \emph {et~al.}}]{mor2017ultrafast}%
     \BibitemOpen
     \bibfield  {author} {\bibinfo {author} {\bibfnamefont {S.}~\bibnamefont
     {Mor}}, \bibinfo {author} {\bibfnamefont {M.}~\bibnamefont {Herzog}},
     \bibinfo {author} {\bibfnamefont {D.}~\bibnamefont {Gole{\v{z}}}}, \bibinfo
     {author} {\bibfnamefont {P.}~\bibnamefont {Werner}}, \bibinfo {author}
     {\bibfnamefont {M.}~\bibnamefont {Eckstein}}, \bibinfo {author}
     {\bibfnamefont {N.}~\bibnamefont {Katayama}}, \bibinfo {author}
     {\bibfnamefont {M.}~\bibnamefont {Nohara}}, \bibinfo {author} {\bibfnamefont
     {H.}~\bibnamefont {Takagi}}, \bibinfo {author} {\bibfnamefont
     {T.}~\bibnamefont {Mizokawa}}, \bibinfo {author} {\bibfnamefont
     {C.}~\bibnamefont {Monney}},  \emph {et~al.},\ }\href@noop {} {\bibfield
     {journal} {\bibinfo  {journal} {Phys. Rev. Lett.}\ }\textbf {\bibinfo
     {volume} {119}},\ \bibinfo {pages} {086401} (\bibinfo {year}
     {2017})}\BibitemShut {NoStop}%
   \bibitem [{\citenamefont {Zhou}\ \emph {et~al.}(2023)\citenamefont {Zhou},
     \citenamefont {Bao}, \citenamefont {Fan}, \citenamefont {Zhou}, \citenamefont
     {Gao}, \citenamefont {Zhong}, \citenamefont {Lin}, \citenamefont {Liu},
     \citenamefont {Yu}, \citenamefont {Tang} \emph
     {et~al.}}]{zhou2023pseudospin}%
     \BibitemOpen
     \bibfield  {author} {\bibinfo {author} {\bibfnamefont {S.}~\bibnamefont
     {Zhou}}, \bibinfo {author} {\bibfnamefont {C.}~\bibnamefont {Bao}}, \bibinfo
     {author} {\bibfnamefont {B.}~\bibnamefont {Fan}}, \bibinfo {author}
     {\bibfnamefont {H.}~\bibnamefont {Zhou}}, \bibinfo {author} {\bibfnamefont
     {Q.}~\bibnamefont {Gao}}, \bibinfo {author} {\bibfnamefont {H.}~\bibnamefont
     {Zhong}}, \bibinfo {author} {\bibfnamefont {T.}~\bibnamefont {Lin}}, \bibinfo
     {author} {\bibfnamefont {H.}~\bibnamefont {Liu}}, \bibinfo {author}
     {\bibfnamefont {P.}~\bibnamefont {Yu}}, \bibinfo {author} {\bibfnamefont
     {P.}~\bibnamefont {Tang}},  \emph {et~al.},\ }\href@noop {} {\bibfield
     {journal} {\bibinfo  {journal} {Nature}\ }\textbf {\bibinfo {volume} {614}},\
     \bibinfo {pages} {75} (\bibinfo {year} {2023})}\BibitemShut {NoStop}%
   \bibitem [{\citenamefont {Liu}\ \emph {et~al.}(2019)\citenamefont {Liu},
     \citenamefont {Ziffer}, \citenamefont {Hansen}, \citenamefont {Wang},\ and\
     \citenamefont {Zhu}}]{liu2019direct}%
     \BibitemOpen
     \bibfield  {author} {\bibinfo {author} {\bibfnamefont {F.}~\bibnamefont
     {Liu}}, \bibinfo {author} {\bibfnamefont {M.~E.}\ \bibnamefont {Ziffer}},
     \bibinfo {author} {\bibfnamefont {K.~R.}\ \bibnamefont {Hansen}}, \bibinfo
     {author} {\bibfnamefont {J.}~\bibnamefont {Wang}}, \ and\ \bibinfo {author}
     {\bibfnamefont {X.}~\bibnamefont {Zhu}},\ }\href@noop {} {\bibfield
     {journal} {\bibinfo  {journal} {Phys. Rev. Lett.}\ }\textbf {\bibinfo
     {volume} {122}},\ \bibinfo {pages} {246803} (\bibinfo {year}
     {2019})}\BibitemShut {NoStop}%
   \bibitem [{\citenamefont {Wegkamp}\ \emph {et~al.}(2014)\citenamefont
     {Wegkamp}, \citenamefont {Herzog}, \citenamefont {Xian}, \citenamefont
     {Gatti}, \citenamefont {Cudazzo}, \citenamefont {McGahan}, \citenamefont
     {Marvel}, \citenamefont {Haglund~Jr}, \citenamefont {Rubio}, \citenamefont
     {Wolf} \emph {et~al.}}]{wegkamp2014instantaneous}%
     \BibitemOpen
     \bibfield  {author} {\bibinfo {author} {\bibfnamefont {D.}~\bibnamefont
     {Wegkamp}}, \bibinfo {author} {\bibfnamefont {M.}~\bibnamefont {Herzog}},
     \bibinfo {author} {\bibfnamefont {L.}~\bibnamefont {Xian}}, \bibinfo {author}
     {\bibfnamefont {M.}~\bibnamefont {Gatti}}, \bibinfo {author} {\bibfnamefont
     {P.}~\bibnamefont {Cudazzo}}, \bibinfo {author} {\bibfnamefont {C.~L.}\
     \bibnamefont {McGahan}}, \bibinfo {author} {\bibfnamefont {R.~E.}\
     \bibnamefont {Marvel}}, \bibinfo {author} {\bibfnamefont {R.~F.}\
     \bibnamefont {Haglund~Jr}}, \bibinfo {author} {\bibfnamefont
     {A.}~\bibnamefont {Rubio}}, \bibinfo {author} {\bibfnamefont
     {M.}~\bibnamefont {Wolf}},  \emph {et~al.},\ }\href@noop {} {\bibfield
     {journal} {\bibinfo  {journal} {Phys. Rev. Lett.}\ }\textbf {\bibinfo
     {volume} {113}},\ \bibinfo {pages} {216401} (\bibinfo {year}
     {2014})}\BibitemShut {NoStop}%
   \bibitem [{\citenamefont {Baldini}\ \emph {et~al.}(2023)\citenamefont
     {Baldini}, \citenamefont {Zong}, \citenamefont {Choi}, \citenamefont {Lee},
     \citenamefont {Michael}, \citenamefont {Windgaetter}, \citenamefont {Mazin},
     \citenamefont {Latini}, \citenamefont {Azoury}, \citenamefont {Lv} \emph
     {et~al.}}]{baldini2023spontaneous}%
     \BibitemOpen
     \bibfield  {author} {\bibinfo {author} {\bibfnamefont {E.}~\bibnamefont
     {Baldini}}, \bibinfo {author} {\bibfnamefont {A.}~\bibnamefont {Zong}},
     \bibinfo {author} {\bibfnamefont {D.}~\bibnamefont {Choi}}, \bibinfo {author}
     {\bibfnamefont {C.}~\bibnamefont {Lee}}, \bibinfo {author} {\bibfnamefont
     {M.~H.}\ \bibnamefont {Michael}}, \bibinfo {author} {\bibfnamefont
     {L.}~\bibnamefont {Windgaetter}}, \bibinfo {author} {\bibfnamefont {I.~I.}\
     \bibnamefont {Mazin}}, \bibinfo {author} {\bibfnamefont {S.}~\bibnamefont
     {Latini}}, \bibinfo {author} {\bibfnamefont {D.}~\bibnamefont {Azoury}},
     \bibinfo {author} {\bibfnamefont {B.}~\bibnamefont {Lv}},  \emph {et~al.},\
     }\href@noop {} {\bibfield  {journal} {\bibinfo  {journal} {Proc. Natl. Acad.
     Sci.}\ }\textbf {\bibinfo {volume} {120}},\ \bibinfo {pages} {e2221688120}
     (\bibinfo {year} {2023})}\BibitemShut {NoStop}%
   \bibitem [{\citenamefont {Andreatta}\ \emph {et~al.}(2019)\citenamefont
     {Andreatta}, \citenamefont {Rostami}, \citenamefont {\ifmmode~\check{C}\else
     \v{C}\fi{}abo}, \citenamefont {Bianchi}, \citenamefont {Sanders},
     \citenamefont {Biswas}, \citenamefont {Cacho}, \citenamefont {Jones},
     \citenamefont {Chapman}, \citenamefont {Springate}, \citenamefont {King},
     \citenamefont {Miwa}, \citenamefont {Balatsky}, \citenamefont {Ulstrup},\
     and\ \citenamefont {Hofmann}}]{andreatta2019}%
     \BibitemOpen
     \bibfield  {author} {\bibinfo {author} {\bibfnamefont {F.}~\bibnamefont
     {Andreatta}}, \bibinfo {author} {\bibfnamefont {H.}~\bibnamefont {Rostami}},
     \bibinfo {author} {\bibfnamefont {A.~G. c. v. a.~c.}\ \bibnamefont
     {\ifmmode~\check{C}\else \v{C}\fi{}abo}}, \bibinfo {author} {\bibfnamefont
     {M.}~\bibnamefont {Bianchi}}, \bibinfo {author} {\bibfnamefont {C.~E.}\
     \bibnamefont {Sanders}}, \bibinfo {author} {\bibfnamefont {D.}~\bibnamefont
     {Biswas}}, \bibinfo {author} {\bibfnamefont {C.}~\bibnamefont {Cacho}},
     \bibinfo {author} {\bibfnamefont {A.~J.~H.}\ \bibnamefont {Jones}}, \bibinfo
     {author} {\bibfnamefont {R.~T.}\ \bibnamefont {Chapman}}, \bibinfo {author}
     {\bibfnamefont {E.}~\bibnamefont {Springate}}, \bibinfo {author}
     {\bibfnamefont {P.~D.~C.}\ \bibnamefont {King}}, \bibinfo {author}
     {\bibfnamefont {J.~A.}\ \bibnamefont {Miwa}}, \bibinfo {author}
     {\bibfnamefont {A.}~\bibnamefont {Balatsky}}, \bibinfo {author}
     {\bibfnamefont {S.}~\bibnamefont {Ulstrup}}, \ and\ \bibinfo {author}
     {\bibfnamefont {P.}~\bibnamefont {Hofmann}},\ }\href {\doibase
     10.1103/PhysRevB.99.165421} {\bibfield  {journal} {\bibinfo  {journal} {Phys.
     Rev. B}\ }\textbf {\bibinfo {volume} {99}},\ \bibinfo {pages} {165421}
     (\bibinfo {year} {2019})}\BibitemShut {NoStop}%
   \bibitem [{\citenamefont {Puppin}\ \emph {et~al.}(2022)\citenamefont {Puppin},
     \citenamefont {Nicholson}, \citenamefont {Monney}, \citenamefont {Deng},
     \citenamefont {Xian}, \citenamefont {Feldl}, \citenamefont {Dong},
     \citenamefont {Dominguez}, \citenamefont {H\"ubener}, \citenamefont {Rubio},
     \citenamefont {Wolf}, \citenamefont {Rettig},\ and\ \citenamefont
     {Ernstorfer}}]{puppin2022}%
     \BibitemOpen
     \bibfield  {author} {\bibinfo {author} {\bibfnamefont {M.}~\bibnamefont
     {Puppin}}, \bibinfo {author} {\bibfnamefont {C.~W.}\ \bibnamefont
     {Nicholson}}, \bibinfo {author} {\bibfnamefont {C.}~\bibnamefont {Monney}},
     \bibinfo {author} {\bibfnamefont {Y.}~\bibnamefont {Deng}}, \bibinfo {author}
     {\bibfnamefont {R.~P.}\ \bibnamefont {Xian}}, \bibinfo {author}
     {\bibfnamefont {J.}~\bibnamefont {Feldl}}, \bibinfo {author} {\bibfnamefont
     {S.}~\bibnamefont {Dong}}, \bibinfo {author} {\bibfnamefont {A.}~\bibnamefont
     {Dominguez}}, \bibinfo {author} {\bibfnamefont {H.}~\bibnamefont
     {H\"ubener}}, \bibinfo {author} {\bibfnamefont {A.}~\bibnamefont {Rubio}},
     \bibinfo {author} {\bibfnamefont {M.}~\bibnamefont {Wolf}}, \bibinfo {author}
     {\bibfnamefont {L.}~\bibnamefont {Rettig}}, \ and\ \bibinfo {author}
     {\bibfnamefont {R.}~\bibnamefont {Ernstorfer}},\ }\href {\doibase
     10.1103/PhysRevB.105.075417} {\bibfield  {journal} {\bibinfo  {journal}
     {Phys. Rev. B}\ }\textbf {\bibinfo {volume} {105}},\ \bibinfo {pages}
     {075417} (\bibinfo {year} {2022})}\BibitemShut {NoStop}%
   \bibitem [{\citenamefont {Mott}(1961)}]{mott1961transition}%
     \BibitemOpen
     \bibfield  {author} {\bibinfo {author} {\bibfnamefont {N.~F.}\ \bibnamefont
     {Mott}},\ }\href@noop {} {\bibfield  {journal} {\bibinfo  {journal} {Philos.
     Mag.}\ }\textbf {\bibinfo {volume} {6}},\ \bibinfo {pages} {287} (\bibinfo
     {year} {1961})}\BibitemShut {NoStop}%
   \bibitem [{\citenamefont {J{\'e}rome}\ \emph {et~al.}(1967)\citenamefont
     {J{\'e}rome}, \citenamefont {Rice},\ and\ \citenamefont
     {Kohn}}]{jerome1967excitonic}%
     \BibitemOpen
     \bibfield  {author} {\bibinfo {author} {\bibfnamefont {D.}~\bibnamefont
     {J{\'e}rome}}, \bibinfo {author} {\bibfnamefont {T.}~\bibnamefont {Rice}}, \
     and\ \bibinfo {author} {\bibfnamefont {W.}~\bibnamefont {Kohn}},\ }\href@noop
     {} {\bibfield  {journal} {\bibinfo  {journal} {Phys. Rev.}\ }\textbf
     {\bibinfo {volume} {158}},\ \bibinfo {pages} {462} (\bibinfo {year}
     {1967})}\BibitemShut {NoStop}%
   \bibitem [{\citenamefont {Halperin}\ and\ \citenamefont
     {Rice}(1968)}]{halperin1968possible}%
     \BibitemOpen
     \bibfield  {author} {\bibinfo {author} {\bibfnamefont {B.}~\bibnamefont
     {Halperin}}\ and\ \bibinfo {author} {\bibfnamefont {T.}~\bibnamefont
     {Rice}},\ }\href@noop {} {\bibfield  {journal} {\bibinfo  {journal} {Rev.
     Mod. Phys.}\ }\textbf {\bibinfo {volume} {40}},\ \bibinfo {pages} {755}
     (\bibinfo {year} {1968})}\BibitemShut {NoStop}%
   \bibitem [{\citenamefont {Eisenstein}(2014)}]{eisenstein2014exciton}%
     \BibitemOpen
     \bibfield  {author} {\bibinfo {author} {\bibfnamefont {J.}~\bibnamefont
     {Eisenstein}},\ }\href@noop {} {\bibfield  {journal} {\bibinfo  {journal}
     {Annu. Rev. Condens. Matter Phys.}\ }\textbf {\bibinfo {volume} {5}},\
     \bibinfo {pages} {159} (\bibinfo {year} {2014})}\BibitemShut {NoStop}%
   \bibitem [{\citenamefont {Li}\ \emph {et~al.}(2017)\citenamefont {Li},
     \citenamefont {Taniguchi}, \citenamefont {Watanabe}, \citenamefont {Hone},\
     and\ \citenamefont {Dean}}]{li2017excitonic}%
     \BibitemOpen
     \bibfield  {author} {\bibinfo {author} {\bibfnamefont {J.}~\bibnamefont
     {Li}}, \bibinfo {author} {\bibfnamefont {T.}~\bibnamefont {Taniguchi}},
     \bibinfo {author} {\bibfnamefont {K.}~\bibnamefont {Watanabe}}, \bibinfo
     {author} {\bibfnamefont {J.}~\bibnamefont {Hone}}, \ and\ \bibinfo {author}
     {\bibfnamefont {C.}~\bibnamefont {Dean}},\ }\href@noop {} {\bibfield
     {journal} {\bibinfo  {journal} {Nat. Phys.}\ }\textbf {\bibinfo {volume}
     {13}},\ \bibinfo {pages} {751} (\bibinfo {year} {2017})}\BibitemShut
     {NoStop}%
   \bibitem [{\citenamefont {Liu}\ \emph {et~al.}(2017)\citenamefont {Liu},
     \citenamefont {Watanabe}, \citenamefont {Taniguchi}, \citenamefont
     {Halperin},\ and\ \citenamefont {Kim}}]{liu2017quantum}%
     \BibitemOpen
     \bibfield  {author} {\bibinfo {author} {\bibfnamefont {X.}~\bibnamefont
     {Liu}}, \bibinfo {author} {\bibfnamefont {K.}~\bibnamefont {Watanabe}},
     \bibinfo {author} {\bibfnamefont {T.}~\bibnamefont {Taniguchi}}, \bibinfo
     {author} {\bibfnamefont {B.~I.}\ \bibnamefont {Halperin}}, \ and\ \bibinfo
     {author} {\bibfnamefont {P.}~\bibnamefont {Kim}},\ }\href@noop {} {\bibfield
     {journal} {\bibinfo  {journal} {Nat. Phys.}\ }\textbf {\bibinfo {volume}
     {13}},\ \bibinfo {pages} {746} (\bibinfo {year} {2017})}\BibitemShut
     {NoStop}%
   \bibitem [{\citenamefont {Cercellier}\ \emph {et~al.}(2007)\citenamefont
     {Cercellier}, \citenamefont {Monney}, \citenamefont {Clerc}, \citenamefont
     {Battaglia}, \citenamefont {Despont}, \citenamefont {Garnier}, \citenamefont
     {Beck}, \citenamefont {Aebi}, \citenamefont {Patthey}, \citenamefont {Berger}
     \emph {et~al.}}]{cercellier2007evidence}%
     \BibitemOpen
     \bibfield  {author} {\bibinfo {author} {\bibfnamefont {H.}~\bibnamefont
     {Cercellier}}, \bibinfo {author} {\bibfnamefont {C.}~\bibnamefont {Monney}},
     \bibinfo {author} {\bibfnamefont {F.}~\bibnamefont {Clerc}}, \bibinfo
     {author} {\bibfnamefont {C.}~\bibnamefont {Battaglia}}, \bibinfo {author}
     {\bibfnamefont {L.}~\bibnamefont {Despont}}, \bibinfo {author} {\bibfnamefont
     {M.}~\bibnamefont {Garnier}}, \bibinfo {author} {\bibfnamefont
     {H.}~\bibnamefont {Beck}}, \bibinfo {author} {\bibfnamefont {P.}~\bibnamefont
     {Aebi}}, \bibinfo {author} {\bibfnamefont {L.}~\bibnamefont {Patthey}},
     \bibinfo {author} {\bibfnamefont {H.}~\bibnamefont {Berger}},  \emph
     {et~al.},\ }\href@noop {} {\bibfield  {journal} {\bibinfo  {journal} {Phys.
     Rev. Lett.}\ }\textbf {\bibinfo {volume} {99}},\ \bibinfo {pages} {146403}
     (\bibinfo {year} {2007})}\BibitemShut {NoStop}%
   \bibitem [{\citenamefont {Monney}\ \emph {et~al.}(2010)\citenamefont {Monney},
     \citenamefont {Schwier}, \citenamefont {Garnier}, \citenamefont {Mariotti},
     \citenamefont {Didiot}, \citenamefont {Beck}, \citenamefont {Aebi},
     \citenamefont {Cercellier}, \citenamefont {Marcus}, \citenamefont
     {Battaglia}, \citenamefont {Berger},\ and\ \citenamefont
     {Titov}}]{monney2010}%
     \BibitemOpen
     \bibfield  {author} {\bibinfo {author} {\bibfnamefont {C.}~\bibnamefont
     {Monney}}, \bibinfo {author} {\bibfnamefont {E.~F.}\ \bibnamefont {Schwier}},
     \bibinfo {author} {\bibfnamefont {M.~G.}\ \bibnamefont {Garnier}}, \bibinfo
     {author} {\bibfnamefont {N.}~\bibnamefont {Mariotti}}, \bibinfo {author}
     {\bibfnamefont {C.}~\bibnamefont {Didiot}}, \bibinfo {author} {\bibfnamefont
     {H.}~\bibnamefont {Beck}}, \bibinfo {author} {\bibfnamefont {P.}~\bibnamefont
     {Aebi}}, \bibinfo {author} {\bibfnamefont {H.}~\bibnamefont {Cercellier}},
     \bibinfo {author} {\bibfnamefont {J.}~\bibnamefont {Marcus}}, \bibinfo
     {author} {\bibfnamefont {C.}~\bibnamefont {Battaglia}}, \bibinfo {author}
     {\bibfnamefont {H.}~\bibnamefont {Berger}}, \ and\ \bibinfo {author}
     {\bibfnamefont {A.~N.}\ \bibnamefont {Titov}},\ }\href {\doibase
     10.1103/PhysRevB.81.155104} {\bibfield  {journal} {\bibinfo  {journal} {Phys.
     Rev. B}\ }\textbf {\bibinfo {volume} {81}},\ \bibinfo {pages} {155104}
     (\bibinfo {year} {2010})}\BibitemShut {NoStop}%
   \bibitem [{\citenamefont {Kogar}\ \emph {et~al.}(2017)\citenamefont {Kogar},
     \citenamefont {Rak}, \citenamefont {Vig}, \citenamefont {Husain},
     \citenamefont {Flicker}, \citenamefont {Joe}, \citenamefont {Venema},
     \citenamefont {MacDougall}, \citenamefont {Chiang}, \citenamefont {Fradkin},
     \citenamefont {van Wezel},\ and\ \citenamefont {Abbamonte}}]{kogar2017}%
     \BibitemOpen
     \bibfield  {author} {\bibinfo {author} {\bibfnamefont {A.}~\bibnamefont
     {Kogar}}, \bibinfo {author} {\bibfnamefont {M.~S.}\ \bibnamefont {Rak}},
     \bibinfo {author} {\bibfnamefont {S.}~\bibnamefont {Vig}}, \bibinfo {author}
     {\bibfnamefont {A.~A.}\ \bibnamefont {Husain}}, \bibinfo {author}
     {\bibfnamefont {F.}~\bibnamefont {Flicker}}, \bibinfo {author} {\bibfnamefont
     {Y.~I.}\ \bibnamefont {Joe}}, \bibinfo {author} {\bibfnamefont
     {L.}~\bibnamefont {Venema}}, \bibinfo {author} {\bibfnamefont {G.~J.}\
     \bibnamefont {MacDougall}}, \bibinfo {author} {\bibfnamefont {T.~C.}\
     \bibnamefont {Chiang}}, \bibinfo {author} {\bibfnamefont {E.}~\bibnamefont
     {Fradkin}}, \bibinfo {author} {\bibfnamefont {J.}~\bibnamefont {van Wezel}},
     \ and\ \bibinfo {author} {\bibfnamefont {P.}~\bibnamefont {Abbamonte}},\
     }\href {\doibase 10.1126/science.aam6432} {\bibfield  {journal} {\bibinfo
     {journal} {Science}\ }\textbf {\bibinfo {volume} {358}},\ \bibinfo {pages}
     {1314} (\bibinfo {year} {2017})}\BibitemShut {NoStop}%
   \bibitem [{\citenamefont {Wakisaka}\ \emph {et~al.}(2009)\citenamefont
     {Wakisaka}, \citenamefont {Sudayama}, \citenamefont {Takubo}, \citenamefont
     {Mizokawa}, \citenamefont {Arita}, \citenamefont {Namatame}, \citenamefont
     {Taniguchi}, \citenamefont {Katayama}, \citenamefont {Nohara},\ and\
     \citenamefont {Takagi}}]{Wakisaka2009}%
     \BibitemOpen
     \bibfield  {author} {\bibinfo {author} {\bibfnamefont {Y.}~\bibnamefont
     {Wakisaka}}, \bibinfo {author} {\bibfnamefont {T.}~\bibnamefont {Sudayama}},
     \bibinfo {author} {\bibfnamefont {K.}~\bibnamefont {Takubo}}, \bibinfo
     {author} {\bibfnamefont {T.}~\bibnamefont {Mizokawa}}, \bibinfo {author}
     {\bibfnamefont {M.}~\bibnamefont {Arita}}, \bibinfo {author} {\bibfnamefont
     {H.}~\bibnamefont {Namatame}}, \bibinfo {author} {\bibfnamefont
     {M.}~\bibnamefont {Taniguchi}}, \bibinfo {author} {\bibfnamefont
     {N.}~\bibnamefont {Katayama}}, \bibinfo {author} {\bibfnamefont
     {M.}~\bibnamefont {Nohara}}, \ and\ \bibinfo {author} {\bibfnamefont
     {H.}~\bibnamefont {Takagi}},\ }\href {\doibase
     10.1103/PhysRevLett.103.026402} {\bibfield  {journal} {\bibinfo  {journal}
     {Phys. Rev. Lett.}\ }\textbf {\bibinfo {volume} {103}},\ \bibinfo {pages}
     {026402} (\bibinfo {year} {2009})}\BibitemShut {NoStop}%
   \bibitem [{\citenamefont {Kaneko}\ \emph {et~al.}(2013)\citenamefont {Kaneko},
     \citenamefont {Toriyama}, \citenamefont {Konishi},\ and\ \citenamefont
     {Ohta}}]{kaneko2013orthorhombic}%
     \BibitemOpen
     \bibfield  {author} {\bibinfo {author} {\bibfnamefont {T.}~\bibnamefont
     {Kaneko}}, \bibinfo {author} {\bibfnamefont {T.}~\bibnamefont {Toriyama}},
     \bibinfo {author} {\bibfnamefont {T.}~\bibnamefont {Konishi}}, \ and\
     \bibinfo {author} {\bibfnamefont {Y.}~\bibnamefont {Ohta}},\ }\href@noop {}
     {\bibfield  {journal} {\bibinfo  {journal} {Phys. Rev. B}\ }\textbf {\bibinfo
     {volume} {87}},\ \bibinfo {pages} {035121} (\bibinfo {year}
     {2013})}\BibitemShut {NoStop}%
   \bibitem [{\citenamefont {Sugimoto}\ \emph {et~al.}(2018)\citenamefont
     {Sugimoto}, \citenamefont {Nishimoto}, \citenamefont {Kaneko},\ and\
     \citenamefont {Ohta}}]{sugimoto2018strong}%
     \BibitemOpen
     \bibfield  {author} {\bibinfo {author} {\bibfnamefont {K.}~\bibnamefont
     {Sugimoto}}, \bibinfo {author} {\bibfnamefont {S.}~\bibnamefont {Nishimoto}},
     \bibinfo {author} {\bibfnamefont {T.}~\bibnamefont {Kaneko}}, \ and\ \bibinfo
     {author} {\bibfnamefont {Y.}~\bibnamefont {Ohta}},\ }\href@noop {} {\bibfield
      {journal} {\bibinfo  {journal} {Phys. Rev. Lett.}\ }\textbf {\bibinfo
     {volume} {120}},\ \bibinfo {pages} {247602} (\bibinfo {year}
     {2018})}\BibitemShut {NoStop}%
   \bibitem [{\citenamefont {Seki}\ \emph {et~al.}(2014)\citenamefont {Seki},
     \citenamefont {Wakisaka}, \citenamefont {Kaneko}, \citenamefont {Toriyama},
     \citenamefont {Konishi}, \citenamefont {Sudayama}, \citenamefont {Saini},
     \citenamefont {Arita}, \citenamefont {Namatame}, \citenamefont {Taniguchi},
     \citenamefont {Katayama}, \citenamefont {Nohara}, \citenamefont {Takagi},
     \citenamefont {Mizokawa},\ and\ \citenamefont {Ohta}}]{seki2014}%
     \BibitemOpen
     \bibfield  {author} {\bibinfo {author} {\bibfnamefont {K.}~\bibnamefont
     {Seki}}, \bibinfo {author} {\bibfnamefont {Y.}~\bibnamefont {Wakisaka}},
     \bibinfo {author} {\bibfnamefont {T.}~\bibnamefont {Kaneko}}, \bibinfo
     {author} {\bibfnamefont {T.}~\bibnamefont {Toriyama}}, \bibinfo {author}
     {\bibfnamefont {T.}~\bibnamefont {Konishi}}, \bibinfo {author} {\bibfnamefont
     {T.}~\bibnamefont {Sudayama}}, \bibinfo {author} {\bibfnamefont {N.~L.}\
     \bibnamefont {Saini}}, \bibinfo {author} {\bibfnamefont {M.}~\bibnamefont
     {Arita}}, \bibinfo {author} {\bibfnamefont {H.}~\bibnamefont {Namatame}},
     \bibinfo {author} {\bibfnamefont {M.}~\bibnamefont {Taniguchi}}, \bibinfo
     {author} {\bibfnamefont {N.}~\bibnamefont {Katayama}}, \bibinfo {author}
     {\bibfnamefont {M.}~\bibnamefont {Nohara}}, \bibinfo {author} {\bibfnamefont
     {H.}~\bibnamefont {Takagi}}, \bibinfo {author} {\bibfnamefont
     {T.}~\bibnamefont {Mizokawa}}, \ and\ \bibinfo {author} {\bibfnamefont
     {Y.}~\bibnamefont {Ohta}},\ }\href {\doibase 10.1103/PhysRevB.90.155116}
     {\bibfield  {journal} {\bibinfo  {journal} {Phys. Rev. B}\ }\textbf {\bibinfo
     {volume} {90}},\ \bibinfo {pages} {155116} (\bibinfo {year}
     {2014})}\BibitemShut {NoStop}%
   \bibitem [{\citenamefont {Lu}\ \emph {et~al.}(2017)\citenamefont {Lu},
     \citenamefont {Kono}, \citenamefont {Larkin}, \citenamefont {Rost},
     \citenamefont {Takayama}, \citenamefont {Boris}, \citenamefont {Keimer},\
     and\ \citenamefont {Takagi}}]{Lu2017}%
     \BibitemOpen
     \bibfield  {author} {\bibinfo {author} {\bibfnamefont {Y.~F.}\ \bibnamefont
     {Lu}}, \bibinfo {author} {\bibfnamefont {H.}~\bibnamefont {Kono}}, \bibinfo
     {author} {\bibfnamefont {T.~I.}\ \bibnamefont {Larkin}}, \bibinfo {author}
     {\bibfnamefont {A.~W.}\ \bibnamefont {Rost}}, \bibinfo {author}
     {\bibfnamefont {T.}~\bibnamefont {Takayama}}, \bibinfo {author}
     {\bibfnamefont {A.~V.}\ \bibnamefont {Boris}}, \bibinfo {author}
     {\bibfnamefont {B.}~\bibnamefont {Keimer}}, \ and\ \bibinfo {author}
     {\bibfnamefont {H.}~\bibnamefont {Takagi}},\ }\href
     {http://dx.doi.org/10.1038/ncomms14408} {\bibfield  {journal} {\bibinfo
     {journal} {Nat. Commun.}\ }\textbf {\bibinfo {volume} {8}} (\bibinfo {year}
     {2017})}\BibitemShut {NoStop}%
   \bibitem [{\citenamefont {Kim}\ \emph {et~al.}(2021)\citenamefont {Kim},
     \citenamefont {Kim}, \citenamefont {Kim}, \citenamefont {Kwon}, \citenamefont
     {Kim},\ and\ \citenamefont {Kim}}]{Kim2021}%
     \BibitemOpen
     \bibfield  {author} {\bibinfo {author} {\bibfnamefont {K.}~\bibnamefont
     {Kim}}, \bibinfo {author} {\bibfnamefont {H.}~\bibnamefont {Kim}}, \bibinfo
     {author} {\bibfnamefont {J.}~\bibnamefont {Kim}}, \bibinfo {author}
     {\bibfnamefont {C.}~\bibnamefont {Kwon}}, \bibinfo {author} {\bibfnamefont
     {J.~S.}\ \bibnamefont {Kim}}, \ and\ \bibinfo {author} {\bibfnamefont
     {B.~J.}\ \bibnamefont {Kim}},\ }\href
     {http://dx.doi.org/10.1038/s41467-021-22133-z} {\bibfield  {journal}
     {\bibinfo  {journal} {Nat. Commun.}\ }\textbf {\bibinfo {volume} {12}}
     (\bibinfo {year} {2021})}\BibitemShut {NoStop}%
   \bibitem [{\citenamefont {Volkov}\ \emph {et~al.}(2021)\citenamefont {Volkov},
     \citenamefont {Ye}, \citenamefont {Lohani}, \citenamefont {Feldman},
     \citenamefont {Kanigel},\ and\ \citenamefont {Blumberg}}]{Volkov2021}%
     \BibitemOpen
     \bibfield  {author} {\bibinfo {author} {\bibfnamefont {P.~A.}\ \bibnamefont
     {Volkov}}, \bibinfo {author} {\bibfnamefont {M.}~\bibnamefont {Ye}}, \bibinfo
     {author} {\bibfnamefont {H.}~\bibnamefont {Lohani}}, \bibinfo {author}
     {\bibfnamefont {I.}~\bibnamefont {Feldman}}, \bibinfo {author} {\bibfnamefont
     {A.}~\bibnamefont {Kanigel}}, \ and\ \bibinfo {author} {\bibfnamefont
     {G.}~\bibnamefont {Blumberg}},\ }\href
     {http://dx.doi.org/10.1038/s41535-021-00351-4} {\bibfield  {journal}
     {\bibinfo  {journal} {npj Quantum Mater.}\ }\textbf {\bibinfo {volume} {6}}
     (\bibinfo {year} {2021})}\BibitemShut {NoStop}%
   \bibitem [{\citenamefont {Ye}\ \emph {et~al.}(2021)\citenamefont {Ye},
     \citenamefont {Volkov}, \citenamefont {Lohani}, \citenamefont {Feldman},
     \citenamefont {Kim}, \citenamefont {Kanigel},\ and\ \citenamefont
     {Blumberg}}]{ye2021}%
     \BibitemOpen
     \bibfield  {author} {\bibinfo {author} {\bibfnamefont {M.}~\bibnamefont
     {Ye}}, \bibinfo {author} {\bibfnamefont {P.~A.}\ \bibnamefont {Volkov}},
     \bibinfo {author} {\bibfnamefont {H.}~\bibnamefont {Lohani}}, \bibinfo
     {author} {\bibfnamefont {I.}~\bibnamefont {Feldman}}, \bibinfo {author}
     {\bibfnamefont {M.}~\bibnamefont {Kim}}, \bibinfo {author} {\bibfnamefont
     {A.}~\bibnamefont {Kanigel}}, \ and\ \bibinfo {author} {\bibfnamefont
     {G.}~\bibnamefont {Blumberg}},\ }\href {\doibase 10.1103/PhysRevB.104.045102}
     {\bibfield  {journal} {\bibinfo  {journal} {Phys. Rev. B}\ }\textbf {\bibinfo
     {volume} {104}},\ \bibinfo {pages} {045102} (\bibinfo {year}
     {2021})}\BibitemShut {NoStop}%
   \bibitem [{\citenamefont {Guan}\ \emph {et~al.}(2023)\citenamefont {Guan},
     \citenamefont {Chen}, \citenamefont {Chen}, \citenamefont {Yao},\ and\
     \citenamefont {Meng}}]{Guan2023}%
     \BibitemOpen
     \bibfield  {author} {\bibinfo {author} {\bibfnamefont {M.}~\bibnamefont
     {Guan}}, \bibinfo {author} {\bibfnamefont {D.}~\bibnamefont {Chen}}, \bibinfo
     {author} {\bibfnamefont {Q.}~\bibnamefont {Chen}}, \bibinfo {author}
     {\bibfnamefont {Y.}~\bibnamefont {Yao}}, \ and\ \bibinfo {author}
     {\bibfnamefont {S.}~\bibnamefont {Meng}},\ }\href {\doibase
     10.1103/PhysRevLett.131.256503} {\bibfield  {journal} {\bibinfo  {journal}
     {Phys. Rev. Lett.}\ }\textbf {\bibinfo {volume} {131}},\ \bibinfo {pages}
     {256503} (\bibinfo {year} {2023})}\BibitemShut {NoStop}%
   \bibitem [{\citenamefont {Murakami}\ \emph {et~al.}(2017)\citenamefont
     {Murakami}, \citenamefont {Gole{\v{z}}}, \citenamefont {Eckstein},\ and\
     \citenamefont {Werner}}]{murakami2017photoinduced}%
     \BibitemOpen
     \bibfield  {author} {\bibinfo {author} {\bibfnamefont {Y.}~\bibnamefont
     {Murakami}}, \bibinfo {author} {\bibfnamefont {D.}~\bibnamefont
     {Gole{\v{z}}}}, \bibinfo {author} {\bibfnamefont {M.}~\bibnamefont
     {Eckstein}}, \ and\ \bibinfo {author} {\bibfnamefont {P.}~\bibnamefont
     {Werner}},\ }\href@noop {} {\bibfield  {journal} {\bibinfo  {journal} {Phys.
     Rev. Lett.}\ }\textbf {\bibinfo {volume} {119}},\ \bibinfo {pages} {247601}
     (\bibinfo {year} {2017})}\BibitemShut {NoStop}%
   \bibitem [{\citenamefont {Gole{\v{z}}}\ \emph {et~al.}(2022)\citenamefont
     {Gole{\v{z}}}, \citenamefont {Dufresne}, \citenamefont {Kim}, \citenamefont
     {Boschini}, \citenamefont {Chu}, \citenamefont {Murakami}, \citenamefont
     {Levy}, \citenamefont {Mills}, \citenamefont {Zhdanovich}, \citenamefont
     {Isobe} \emph {et~al.}}]{golevz2022unveiling}%
     \BibitemOpen
     \bibfield  {author} {\bibinfo {author} {\bibfnamefont {D.}~\bibnamefont
     {Gole{\v{z}}}}, \bibinfo {author} {\bibfnamefont {S.~K.}\ \bibnamefont
     {Dufresne}}, \bibinfo {author} {\bibfnamefont {M.-J.}\ \bibnamefont {Kim}},
     \bibinfo {author} {\bibfnamefont {F.}~\bibnamefont {Boschini}}, \bibinfo
     {author} {\bibfnamefont {H.}~\bibnamefont {Chu}}, \bibinfo {author}
     {\bibfnamefont {Y.}~\bibnamefont {Murakami}}, \bibinfo {author}
     {\bibfnamefont {G.}~\bibnamefont {Levy}}, \bibinfo {author} {\bibfnamefont
     {A.~K.}\ \bibnamefont {Mills}}, \bibinfo {author} {\bibfnamefont
     {S.}~\bibnamefont {Zhdanovich}}, \bibinfo {author} {\bibfnamefont
     {M.}~\bibnamefont {Isobe}},  \emph {et~al.},\ }\href@noop {} {\bibfield
     {journal} {\bibinfo  {journal} {Phys. Rev. B}\ }\textbf {\bibinfo {volume}
     {106}},\ \bibinfo {pages} {L121106} (\bibinfo {year} {2022})}\BibitemShut
     {NoStop}%
   \bibitem [{\citenamefont {Saha}\ \emph {et~al.}(2021)\citenamefont {Saha},
     \citenamefont {Gole{\v{z}}}, \citenamefont {De~Ninno}, \citenamefont
     {Mravlje}, \citenamefont {Murakami}, \citenamefont {Ressel}, \citenamefont
     {Stupar},\ and\ \citenamefont {Ribi{\v{c}}}}]{saha2021photoinduced}%
     \BibitemOpen
     \bibfield  {author} {\bibinfo {author} {\bibfnamefont {T.}~\bibnamefont
     {Saha}}, \bibinfo {author} {\bibfnamefont {D.}~\bibnamefont {Gole{\v{z}}}},
     \bibinfo {author} {\bibfnamefont {G.}~\bibnamefont {De~Ninno}}, \bibinfo
     {author} {\bibfnamefont {J.}~\bibnamefont {Mravlje}}, \bibinfo {author}
     {\bibfnamefont {Y.}~\bibnamefont {Murakami}}, \bibinfo {author}
     {\bibfnamefont {B.}~\bibnamefont {Ressel}}, \bibinfo {author} {\bibfnamefont
     {M.}~\bibnamefont {Stupar}}, \ and\ \bibinfo {author} {\bibfnamefont {P.~R.}\
     \bibnamefont {Ribi{\v{c}}}},\ }\href@noop {} {\bibfield  {journal} {\bibinfo
     {journal} {Phys. Rev. B}\ }\textbf {\bibinfo {volume} {103}},\ \bibinfo
     {pages} {144304} (\bibinfo {year} {2021})}\BibitemShut {NoStop}%
   \bibitem [{\citenamefont {Tang}\ \emph
     {et~al.}(2020{\natexlab{a}})\citenamefont {Tang}, \citenamefont {Wang},
     \citenamefont {Duan}, \citenamefont {Yang}, \citenamefont {Huang},
     \citenamefont {Guo}, \citenamefont {Qian},\ and\ \citenamefont
     {Zhang}}]{tang2020non}%
     \BibitemOpen
     \bibfield  {author} {\bibinfo {author} {\bibfnamefont {T.}~\bibnamefont
     {Tang}}, \bibinfo {author} {\bibfnamefont {H.}~\bibnamefont {Wang}}, \bibinfo
     {author} {\bibfnamefont {S.}~\bibnamefont {Duan}}, \bibinfo {author}
     {\bibfnamefont {Y.}~\bibnamefont {Yang}}, \bibinfo {author} {\bibfnamefont
     {C.}~\bibnamefont {Huang}}, \bibinfo {author} {\bibfnamefont
     {Y.}~\bibnamefont {Guo}}, \bibinfo {author} {\bibfnamefont {D.}~\bibnamefont
     {Qian}}, \ and\ \bibinfo {author} {\bibfnamefont {W.}~\bibnamefont {Zhang}},\
     }\href@noop {} {\bibfield  {journal} {\bibinfo  {journal} {Phys. Rev. B}\
     }\textbf {\bibinfo {volume} {101}},\ \bibinfo {pages} {235148} (\bibinfo
     {year} {2020}{\natexlab{a}})}\BibitemShut {NoStop}%
   \bibitem [{\citenamefont {Mazza}\ \emph {et~al.}(2020)\citenamefont {Mazza},
     \citenamefont {R{\"o}sner}, \citenamefont {Windg{\"a}tter}, \citenamefont
     {Latini}, \citenamefont {H{\"u}bener}, \citenamefont {Millis}, \citenamefont
     {Rubio},\ and\ \citenamefont {Georges}}]{mazza2020nature}%
     \BibitemOpen
     \bibfield  {author} {\bibinfo {author} {\bibfnamefont {G.}~\bibnamefont
     {Mazza}}, \bibinfo {author} {\bibfnamefont {M.}~\bibnamefont {R{\"o}sner}},
     \bibinfo {author} {\bibfnamefont {L.}~\bibnamefont {Windg{\"a}tter}},
     \bibinfo {author} {\bibfnamefont {S.}~\bibnamefont {Latini}}, \bibinfo
     {author} {\bibfnamefont {H.}~\bibnamefont {H{\"u}bener}}, \bibinfo {author}
     {\bibfnamefont {A.~J.}\ \bibnamefont {Millis}}, \bibinfo {author}
     {\bibfnamefont {A.}~\bibnamefont {Rubio}}, \ and\ \bibinfo {author}
     {\bibfnamefont {A.}~\bibnamefont {Georges}},\ }\href@noop {} {\bibfield
     {journal} {\bibinfo  {journal} {Phys. Rev. Lett.}\ }\textbf {\bibinfo
     {volume} {124}},\ \bibinfo {pages} {197601} (\bibinfo {year}
     {2020})}\BibitemShut {NoStop}%
   \bibitem [{\citenamefont {Gole{\v{z}}}\ \emph {et~al.}(2016)\citenamefont
     {Gole{\v{z}}}, \citenamefont {Werner},\ and\ \citenamefont
     {Eckstein}}]{golevz2016photoinduced}%
     \BibitemOpen
     \bibfield  {author} {\bibinfo {author} {\bibfnamefont {D.}~\bibnamefont
     {Gole{\v{z}}}}, \bibinfo {author} {\bibfnamefont {P.}~\bibnamefont {Werner}},
     \ and\ \bibinfo {author} {\bibfnamefont {M.}~\bibnamefont {Eckstein}},\
     }\href@noop {} {\bibfield  {journal} {\bibinfo  {journal} {Phys. Rev. B}\
     }\textbf {\bibinfo {volume} {94}},\ \bibinfo {pages} {035121} (\bibinfo
     {year} {2016})}\BibitemShut {NoStop}%
   \bibitem [{\citenamefont {Sch{\"u}ler}\ \emph {et~al.}(2020)\citenamefont
     {Sch{\"u}ler}, \citenamefont {Gole{\v{z}}}, \citenamefont {Murakami},
     \citenamefont {Bittner}, \citenamefont {Herrmann}, \citenamefont {Strand},
     \citenamefont {Werner},\ and\ \citenamefont {Eckstein}}]{schuler2020nessi}%
     \BibitemOpen
     \bibfield  {author} {\bibinfo {author} {\bibfnamefont {M.}~\bibnamefont
     {Sch{\"u}ler}}, \bibinfo {author} {\bibfnamefont {D.}~\bibnamefont
     {Gole{\v{z}}}}, \bibinfo {author} {\bibfnamefont {Y.}~\bibnamefont
     {Murakami}}, \bibinfo {author} {\bibfnamefont {N.}~\bibnamefont {Bittner}},
     \bibinfo {author} {\bibfnamefont {A.}~\bibnamefont {Herrmann}}, \bibinfo
     {author} {\bibfnamefont {H.~U.}\ \bibnamefont {Strand}}, \bibinfo {author}
     {\bibfnamefont {P.}~\bibnamefont {Werner}}, \ and\ \bibinfo {author}
     {\bibfnamefont {M.}~\bibnamefont {Eckstein}},\ }\href@noop {} {\bibfield
     {journal} {\bibinfo  {journal} {Comput. Phys. Commun.}\ }\textbf {\bibinfo
     {volume} {257}},\ \bibinfo {pages} {107484} (\bibinfo {year}
     {2020})}\BibitemShut {NoStop}%
   \bibitem [{\citenamefont {Keldysh}\ \emph {et~al.}(1965)\citenamefont {Keldysh}
     \emph {et~al.}}]{keldysh1965diagram}%
     \BibitemOpen
     \bibfield  {author} {\bibinfo {author} {\bibfnamefont {L.~V.}\ \bibnamefont
     {Keldysh}} \emph {et~al.},\ }\href@noop {} {\bibfield  {journal} {\bibinfo
     {journal} {Sov. Phys. JETP}\ }\textbf {\bibinfo {volume} {20}},\ \bibinfo
     {pages} {1018} (\bibinfo {year} {1965})}\BibitemShut {NoStop}%
   \bibitem [{\citenamefont {Kadanoff}\ and\ \citenamefont
     {Baym}(1962)}]{kadanoff1962quantum}%
     \BibitemOpen
     \bibfield  {author} {\bibinfo {author} {\bibfnamefont {L.~P.}\ \bibnamefont
     {Kadanoff}}\ and\ \bibinfo {author} {\bibfnamefont {G.}~\bibnamefont
     {Baym}},\ }\href@noop {} {\emph {\bibinfo {title} {Quantum statistical
     mechanics}}}\ (\bibinfo  {publisher} {Benjamin, New York},\ \bibinfo {year}
     {1962})\BibitemShut {NoStop}%
   \bibitem [{SM()}]{SM}%
     \BibitemOpen
     \href@noop {} {\bibinfo  {journal} {See Supplemental Material for details}\
     }\BibitemShut {NoStop}%
   \bibitem [{\citenamefont {Gole{\v{z}}}\ \emph {et~al.}(2019)\citenamefont
     {Gole{\v{z}}}, \citenamefont {Eckstein},\ and\ \citenamefont
     {Werner}}]{golevz2019multiband}%
     \BibitemOpen
   \bibfield  {journal} {  }\bibfield  {author} {\bibinfo {author} {\bibfnamefont
     {D.}~\bibnamefont {Gole{\v{z}}}}, \bibinfo {author} {\bibfnamefont
     {M.}~\bibnamefont {Eckstein}}, \ and\ \bibinfo {author} {\bibfnamefont
     {P.}~\bibnamefont {Werner}},\ }\href@noop {} {\bibfield  {journal} {\bibinfo
     {journal} {Phys. Rev. B}\ }\textbf {\bibinfo {volume} {100}},\ \bibinfo
     {pages} {235117} (\bibinfo {year} {2019})}\BibitemShut {NoStop}%
   \bibitem [{\citenamefont {Yue}\ and\ \citenamefont
     {Gaarde}(2020)}]{yue2020structure}%
     \BibitemOpen
     \bibfield  {author} {\bibinfo {author} {\bibfnamefont {L.}~\bibnamefont
     {Yue}}\ and\ \bibinfo {author} {\bibfnamefont {M.~B.}\ \bibnamefont
     {Gaarde}},\ }\href@noop {} {\bibfield  {journal} {\bibinfo  {journal} {Phys.
     Rev. A}\ }\textbf {\bibinfo {volume} {101}},\ \bibinfo {pages} {053411}
     (\bibinfo {year} {2020})}\BibitemShut {NoStop}%
   \bibitem [{\citenamefont {Silva}\ \emph {et~al.}(2019)\citenamefont {Silva},
     \citenamefont {Mart{\'\i}n},\ and\ \citenamefont {Ivanov}}]{silva2019high}%
     \BibitemOpen
     \bibfield  {author} {\bibinfo {author} {\bibfnamefont {R.}~\bibnamefont
     {Silva}}, \bibinfo {author} {\bibfnamefont {F.}~\bibnamefont {Mart{\'\i}n}},
     \ and\ \bibinfo {author} {\bibfnamefont {M.}~\bibnamefont {Ivanov}},\
     }\href@noop {} {\bibfield  {journal} {\bibinfo  {journal} {Phys. Rev. B}\
     }\textbf {\bibinfo {volume} {100}},\ \bibinfo {pages} {195201} (\bibinfo
     {year} {2019})}\BibitemShut {NoStop}%
   \bibitem [{\citenamefont {Li}\ \emph {et~al.}(2020)\citenamefont {Li},
     \citenamefont {Golez}, \citenamefont {Mazza}, \citenamefont {Millis},
     \citenamefont {Georges},\ and\ \citenamefont
     {Eckstein}}]{li2020electromagnetic}%
     \BibitemOpen
     \bibfield  {author} {\bibinfo {author} {\bibfnamefont {J.}~\bibnamefont
     {Li}}, \bibinfo {author} {\bibfnamefont {D.}~\bibnamefont {Golez}}, \bibinfo
     {author} {\bibfnamefont {G.}~\bibnamefont {Mazza}}, \bibinfo {author}
     {\bibfnamefont {A.~J.}\ \bibnamefont {Millis}}, \bibinfo {author}
     {\bibfnamefont {A.}~\bibnamefont {Georges}}, \ and\ \bibinfo {author}
     {\bibfnamefont {M.}~\bibnamefont {Eckstein}},\ }\href@noop {} {\bibfield
     {journal} {\bibinfo  {journal} {Phys. Rev. B}\ }\textbf {\bibinfo {volume}
     {101}},\ \bibinfo {pages} {205140} (\bibinfo {year} {2020})}\BibitemShut
     {NoStop}%
   \bibitem [{\citenamefont {Mor}\ \emph {et~al.}(2022)\citenamefont {Mor},
     \citenamefont {Herzog}, \citenamefont {Monney},\ and\ \citenamefont
     {St{\"a}hler}}]{mor2022ultrafast}%
     \BibitemOpen
     \bibfield  {author} {\bibinfo {author} {\bibfnamefont {S.}~\bibnamefont
     {Mor}}, \bibinfo {author} {\bibfnamefont {M.}~\bibnamefont {Herzog}},
     \bibinfo {author} {\bibfnamefont {C.}~\bibnamefont {Monney}}, \ and\ \bibinfo
     {author} {\bibfnamefont {J.}~\bibnamefont {St{\"a}hler}},\ }\href@noop {}
     {\bibfield  {journal} {\bibinfo  {journal} {Prog. Surf. Sci.}\ ,\ \bibinfo
     {pages} {100679}} (\bibinfo {year} {2022})}\BibitemShut {NoStop}%
   \bibitem [{\citenamefont {Cilento}\ \emph {et~al.}(2018)\citenamefont
     {Cilento}, \citenamefont {Manzoni}, \citenamefont {Sterzi}, \citenamefont
     {Peli}, \citenamefont {Ronchi}, \citenamefont {Crepaldi}, \citenamefont
     {Boschini}, \citenamefont {Cacho}, \citenamefont {Chapman}, \citenamefont
     {Springate} \emph {et~al.}}]{cilento2018dynamics}%
     \BibitemOpen
     \bibfield  {author} {\bibinfo {author} {\bibfnamefont {F.}~\bibnamefont
     {Cilento}}, \bibinfo {author} {\bibfnamefont {G.}~\bibnamefont {Manzoni}},
     \bibinfo {author} {\bibfnamefont {A.}~\bibnamefont {Sterzi}}, \bibinfo
     {author} {\bibfnamefont {S.}~\bibnamefont {Peli}}, \bibinfo {author}
     {\bibfnamefont {A.}~\bibnamefont {Ronchi}}, \bibinfo {author} {\bibfnamefont
     {A.}~\bibnamefont {Crepaldi}}, \bibinfo {author} {\bibfnamefont
     {F.}~\bibnamefont {Boschini}}, \bibinfo {author} {\bibfnamefont
     {C.}~\bibnamefont {Cacho}}, \bibinfo {author} {\bibfnamefont
     {R.}~\bibnamefont {Chapman}}, \bibinfo {author} {\bibfnamefont
     {E.}~\bibnamefont {Springate}},  \emph {et~al.},\ }\href@noop {} {\bibfield
     {journal} {\bibinfo  {journal} {Sci. Adv.}\ }\textbf {\bibinfo {volume}
     {4}},\ \bibinfo {pages} {eaar1998} (\bibinfo {year} {2018})}\BibitemShut
     {NoStop}%
   \bibitem [{\citenamefont {Ma}\ \emph {et~al.}(2022)\citenamefont {Ma},
     \citenamefont {Wang}, \citenamefont {Mao}, \citenamefont {Yuan},
     \citenamefont {Yu}, \citenamefont {Liu}, \citenamefont {Peng}, \citenamefont
     {Zheng},\ and\ \citenamefont {Yin}}]{ma2022ta}%
     \BibitemOpen
     \bibfield  {author} {\bibinfo {author} {\bibfnamefont {X.}~\bibnamefont
     {Ma}}, \bibinfo {author} {\bibfnamefont {G.}~\bibnamefont {Wang}}, \bibinfo
     {author} {\bibfnamefont {H.}~\bibnamefont {Mao}}, \bibinfo {author}
     {\bibfnamefont {Z.}~\bibnamefont {Yuan}}, \bibinfo {author} {\bibfnamefont
     {T.}~\bibnamefont {Yu}}, \bibinfo {author} {\bibfnamefont {R.}~\bibnamefont
     {Liu}}, \bibinfo {author} {\bibfnamefont {Y.}~\bibnamefont {Peng}}, \bibinfo
     {author} {\bibfnamefont {P.}~\bibnamefont {Zheng}}, \ and\ \bibinfo {author}
     {\bibfnamefont {Z.}~\bibnamefont {Yin}},\ }\href@noop {} {\bibfield
     {journal} {\bibinfo  {journal} {Phys. Rev. B}\ }\textbf {\bibinfo {volume}
     {105}},\ \bibinfo {pages} {035138} (\bibinfo {year} {2022})}\BibitemShut
     {NoStop}%
   \bibitem [{\citenamefont {Spataru}\ \emph {et~al.}(2004)\citenamefont
     {Spataru}, \citenamefont {Benedict},\ and\ \citenamefont
     {Louie}}]{spataru2004ab}%
     \BibitemOpen
     \bibfield  {author} {\bibinfo {author} {\bibfnamefont {C.~D.}\ \bibnamefont
     {Spataru}}, \bibinfo {author} {\bibfnamefont {L.~X.}\ \bibnamefont
     {Benedict}}, \ and\ \bibinfo {author} {\bibfnamefont {S.~G.}\ \bibnamefont
     {Louie}},\ }\href@noop {} {\bibfield  {journal} {\bibinfo  {journal} {Phys.
     Rev. B}\ }\textbf {\bibinfo {volume} {69}},\ \bibinfo {pages} {205204}
     (\bibinfo {year} {2004})}\BibitemShut {NoStop}%
   \bibitem [{\citenamefont {Hu}\ \emph {et~al.}(2022)\citenamefont {Hu},
     \citenamefont {Zhao}, \citenamefont {Lian}, \citenamefont {Liu},
     \citenamefont {Guan},\ and\ \citenamefont {Meng}}]{hu2022tracking}%
     \BibitemOpen
     \bibfield  {author} {\bibinfo {author} {\bibfnamefont {S.-Q.}\ \bibnamefont
     {Hu}}, \bibinfo {author} {\bibfnamefont {H.}~\bibnamefont {Zhao}}, \bibinfo
     {author} {\bibfnamefont {C.}~\bibnamefont {Lian}}, \bibinfo {author}
     {\bibfnamefont {X.-B.}\ \bibnamefont {Liu}}, \bibinfo {author} {\bibfnamefont
     {M.-X.}\ \bibnamefont {Guan}}, \ and\ \bibinfo {author} {\bibfnamefont
     {S.}~\bibnamefont {Meng}},\ }\href@noop {} {\bibfield  {journal} {\bibinfo
     {journal} {npj Quantum Mater.}\ }\textbf {\bibinfo {volume} {7}},\ \bibinfo
     {pages} {14} (\bibinfo {year} {2022})}\BibitemShut {NoStop}%
   \bibitem [{\citenamefont {Mor}\ \emph {et~al.}(2018)\citenamefont {Mor},
     \citenamefont {Herzog}, \citenamefont {Noack}, \citenamefont {Katayama},
     \citenamefont {Nohara}, \citenamefont {Takagi}, \citenamefont {Trunschke},
     \citenamefont {Mizokawa}, \citenamefont {Monney},\ and\ \citenamefont
     {St\"ahler}}]{mor2018}%
     \BibitemOpen
     \bibfield  {author} {\bibinfo {author} {\bibfnamefont {S.}~\bibnamefont
     {Mor}}, \bibinfo {author} {\bibfnamefont {M.}~\bibnamefont {Herzog}},
     \bibinfo {author} {\bibfnamefont {J.}~\bibnamefont {Noack}}, \bibinfo
     {author} {\bibfnamefont {N.}~\bibnamefont {Katayama}}, \bibinfo {author}
     {\bibfnamefont {M.}~\bibnamefont {Nohara}}, \bibinfo {author} {\bibfnamefont
     {H.}~\bibnamefont {Takagi}}, \bibinfo {author} {\bibfnamefont
     {A.}~\bibnamefont {Trunschke}}, \bibinfo {author} {\bibfnamefont
     {T.}~\bibnamefont {Mizokawa}}, \bibinfo {author} {\bibfnamefont
     {C.}~\bibnamefont {Monney}}, \ and\ \bibinfo {author} {\bibfnamefont
     {J.}~\bibnamefont {St\"ahler}},\ }\href {\doibase 10.1103/PhysRevB.97.115154}
     {\bibfield  {journal} {\bibinfo  {journal} {Phys. Rev. B}\ }\textbf {\bibinfo
     {volume} {97}},\ \bibinfo {pages} {115154} (\bibinfo {year}
     {2018})}\BibitemShut {NoStop}%
   \bibitem [{\citenamefont {Jiang}\ \emph {et~al.}(2018)\citenamefont {Jiang},
     \citenamefont {Chen}, \citenamefont {Zheng}, \citenamefont {Xu},\ and\
     \citenamefont {Tang}}]{jiang2018photo}%
     \BibitemOpen
     \bibfield  {author} {\bibinfo {author} {\bibfnamefont {T.}~\bibnamefont
     {Jiang}}, \bibinfo {author} {\bibfnamefont {R.}~\bibnamefont {Chen}},
     \bibinfo {author} {\bibfnamefont {X.}~\bibnamefont {Zheng}}, \bibinfo
     {author} {\bibfnamefont {Z.}~\bibnamefont {Xu}}, \ and\ \bibinfo {author}
     {\bibfnamefont {Y.}~\bibnamefont {Tang}},\ }\href@noop {} {\bibfield
     {journal} {\bibinfo  {journal} {Opt. Express}\ }\textbf {\bibinfo {volume}
     {26}},\ \bibinfo {pages} {859} (\bibinfo {year} {2018})}\BibitemShut
     {NoStop}%
   \bibitem [{\citenamefont {Bera}\ \emph {et~al.}(2021)\citenamefont {Bera},
     \citenamefont {Shrivastava}, \citenamefont {Bramhachari}, \citenamefont
     {Zhang}, \citenamefont {Poonia}, \citenamefont {Mandal}, \citenamefont
     {Miller}, \citenamefont {Beard}, \citenamefont {Agarwal},\ and\ \citenamefont
     {Adarsh}}]{bera2021atomlike}%
     \BibitemOpen
     \bibfield  {author} {\bibinfo {author} {\bibfnamefont {S.~K.}\ \bibnamefont
     {Bera}}, \bibinfo {author} {\bibfnamefont {M.}~\bibnamefont {Shrivastava}},
     \bibinfo {author} {\bibfnamefont {K.}~\bibnamefont {Bramhachari}}, \bibinfo
     {author} {\bibfnamefont {H.}~\bibnamefont {Zhang}}, \bibinfo {author}
     {\bibfnamefont {A.~K.}\ \bibnamefont {Poonia}}, \bibinfo {author}
     {\bibfnamefont {D.}~\bibnamefont {Mandal}}, \bibinfo {author} {\bibfnamefont
     {E.~M.}\ \bibnamefont {Miller}}, \bibinfo {author} {\bibfnamefont {M.~C.}\
     \bibnamefont {Beard}}, \bibinfo {author} {\bibfnamefont {A.}~\bibnamefont
     {Agarwal}}, \ and\ \bibinfo {author} {\bibfnamefont {K.}~\bibnamefont
     {Adarsh}},\ }\href@noop {} {\bibfield  {journal} {\bibinfo  {journal} {Phys.
     Rev. B}\ }\textbf {\bibinfo {volume} {104}},\ \bibinfo {pages} {L201404}
     (\bibinfo {year} {2021})}\BibitemShut {NoStop}%
   \bibitem [{\citenamefont {Pagliara}\ \emph {et~al.}(2011)\citenamefont
     {Pagliara}, \citenamefont {Galimberti}, \citenamefont {Mor}, \citenamefont
     {Montagnese}, \citenamefont {Ferrini}, \citenamefont {Grandi}, \citenamefont
     {Galinetto},\ and\ \citenamefont {Parmigiani}}]{pagliara2011photoinduced}%
     \BibitemOpen
     \bibfield  {author} {\bibinfo {author} {\bibfnamefont {S.}~\bibnamefont
     {Pagliara}}, \bibinfo {author} {\bibfnamefont {G.}~\bibnamefont
     {Galimberti}}, \bibinfo {author} {\bibfnamefont {S.}~\bibnamefont {Mor}},
     \bibinfo {author} {\bibfnamefont {M.}~\bibnamefont {Montagnese}}, \bibinfo
     {author} {\bibfnamefont {G.}~\bibnamefont {Ferrini}}, \bibinfo {author}
     {\bibfnamefont {M.}~\bibnamefont {Grandi}}, \bibinfo {author} {\bibfnamefont
     {P.}~\bibnamefont {Galinetto}}, \ and\ \bibinfo {author} {\bibfnamefont
     {F.}~\bibnamefont {Parmigiani}},\ }\href@noop {} {\bibfield  {journal}
     {\bibinfo  {journal} {J. Am. Chem. Soc.}\ }\textbf {\bibinfo {volume}
     {133}},\ \bibinfo {pages} {6318} (\bibinfo {year} {2011})}\BibitemShut
     {NoStop}%
   \bibitem [{\citenamefont {Ligges}\ \emph
     {et~al.}(2018{\natexlab{b}})\citenamefont {Ligges}, \citenamefont {Avigo},
     \citenamefont {Gole{\v{z}}}, \citenamefont {Strand}, \citenamefont {Beyazit},
     \citenamefont {Hanff}, \citenamefont {Diekmann}, \citenamefont {Stojchevska},
     \citenamefont {Kall{\"a}ne}, \citenamefont {Zhou} \emph
     {et~al.}}]{ligges2018ultrafast}%
     \BibitemOpen
     \bibfield  {author} {\bibinfo {author} {\bibfnamefont {M.}~\bibnamefont
     {Ligges}}, \bibinfo {author} {\bibfnamefont {I.}~\bibnamefont {Avigo}},
     \bibinfo {author} {\bibfnamefont {D.}~\bibnamefont {Gole{\v{z}}}}, \bibinfo
     {author} {\bibfnamefont {H.~U.}\ \bibnamefont {Strand}}, \bibinfo {author}
     {\bibfnamefont {Y.}~\bibnamefont {Beyazit}}, \bibinfo {author} {\bibfnamefont
     {K.}~\bibnamefont {Hanff}}, \bibinfo {author} {\bibfnamefont
     {F.}~\bibnamefont {Diekmann}}, \bibinfo {author} {\bibfnamefont
     {L.}~\bibnamefont {Stojchevska}}, \bibinfo {author} {\bibfnamefont
     {M.}~\bibnamefont {Kall{\"a}ne}}, \bibinfo {author} {\bibfnamefont
     {P.}~\bibnamefont {Zhou}},  \emph {et~al.},\ }\href@noop {} {\bibfield
     {journal} {\bibinfo  {journal} {Phys. Rev. Lett.}\ }\textbf {\bibinfo
     {volume} {120}},\ \bibinfo {pages} {166401} (\bibinfo {year}
     {2018}{\natexlab{b}})}\BibitemShut {NoStop}%
   \bibitem [{\citenamefont {Li}\ \emph {et~al.}(2021)\citenamefont {Li},
     \citenamefont {Li}, \citenamefont {Naik}, \citenamefont {Xie}, \citenamefont
     {Li}, \citenamefont {Wang}, \citenamefont {Regan}, \citenamefont {Wang},
     \citenamefont {Zhao}, \citenamefont {Zhao}, \citenamefont {Kahn},
     \citenamefont {Yumigeta}, \citenamefont {Blei}, \citenamefont {Taniguchi},
     \citenamefont {Watanabe}, \citenamefont {Tongay}, \citenamefont {Zettl},
     \citenamefont {Louie}, \citenamefont {Wang},\ and\ \citenamefont
     {Crommie}}]{Li2021}%
     \BibitemOpen
     \bibfield  {author} {\bibinfo {author} {\bibfnamefont {H.}~\bibnamefont
     {Li}}, \bibinfo {author} {\bibfnamefont {S.}~\bibnamefont {Li}}, \bibinfo
     {author} {\bibfnamefont {M.~H.}\ \bibnamefont {Naik}}, \bibinfo {author}
     {\bibfnamefont {J.}~\bibnamefont {Xie}}, \bibinfo {author} {\bibfnamefont
     {X.}~\bibnamefont {Li}}, \bibinfo {author} {\bibfnamefont {J.}~\bibnamefont
     {Wang}}, \bibinfo {author} {\bibfnamefont {E.}~\bibnamefont {Regan}},
     \bibinfo {author} {\bibfnamefont {D.}~\bibnamefont {Wang}}, \bibinfo {author}
     {\bibfnamefont {W.}~\bibnamefont {Zhao}}, \bibinfo {author} {\bibfnamefont
     {S.}~\bibnamefont {Zhao}}, \bibinfo {author} {\bibfnamefont {S.}~\bibnamefont
     {Kahn}}, \bibinfo {author} {\bibfnamefont {K.}~\bibnamefont {Yumigeta}},
     \bibinfo {author} {\bibfnamefont {M.}~\bibnamefont {Blei}}, \bibinfo {author}
     {\bibfnamefont {T.}~\bibnamefont {Taniguchi}}, \bibinfo {author}
     {\bibfnamefont {K.}~\bibnamefont {Watanabe}}, \bibinfo {author}
     {\bibfnamefont {S.}~\bibnamefont {Tongay}}, \bibinfo {author} {\bibfnamefont
     {A.}~\bibnamefont {Zettl}}, \bibinfo {author} {\bibfnamefont {S.~G.}\
     \bibnamefont {Louie}}, \bibinfo {author} {\bibfnamefont {F.}~\bibnamefont
     {Wang}}, \ and\ \bibinfo {author} {\bibfnamefont {M.~F.}\ \bibnamefont
     {Crommie}},\ }\href {\doibase 10.1038/s41563-021-00923-6} {\bibfield
     {journal} {\bibinfo  {journal} {Nature Materials}\ }\textbf {\bibinfo
     {volume} {20}},\ \bibinfo {pages} {945–950} (\bibinfo {year}
     {2021})}\BibitemShut {NoStop}%
   \bibitem [{\citenamefont {Tang}\ \emph
     {et~al.}(2020{\natexlab{b}})\citenamefont {Tang}, \citenamefont {Li},
     \citenamefont {Li}, \citenamefont {Xu}, \citenamefont {Liu}, \citenamefont
     {Barmak}, \citenamefont {Watanabe}, \citenamefont {Taniguchi}, \citenamefont
     {MacDonald}, \citenamefont {Shan},\ and\ \citenamefont {Mak}}]{Tang2020}%
     \BibitemOpen
     \bibfield  {author} {\bibinfo {author} {\bibfnamefont {Y.}~\bibnamefont
     {Tang}}, \bibinfo {author} {\bibfnamefont {L.}~\bibnamefont {Li}}, \bibinfo
     {author} {\bibfnamefont {T.}~\bibnamefont {Li}}, \bibinfo {author}
     {\bibfnamefont {Y.}~\bibnamefont {Xu}}, \bibinfo {author} {\bibfnamefont
     {S.}~\bibnamefont {Liu}}, \bibinfo {author} {\bibfnamefont {K.}~\bibnamefont
     {Barmak}}, \bibinfo {author} {\bibfnamefont {K.}~\bibnamefont {Watanabe}},
     \bibinfo {author} {\bibfnamefont {T.}~\bibnamefont {Taniguchi}}, \bibinfo
     {author} {\bibfnamefont {A.~H.}\ \bibnamefont {MacDonald}}, \bibinfo {author}
     {\bibfnamefont {J.}~\bibnamefont {Shan}}, \ and\ \bibinfo {author}
     {\bibfnamefont {K.~F.}\ \bibnamefont {Mak}},\ }\href {\doibase
     10.1038/s41586-020-2085-3} {\bibfield  {journal} {\bibinfo  {journal}
     {Nature}\ }\textbf {\bibinfo {volume} {579}},\ \bibinfo {pages} {353–358}
     (\bibinfo {year} {2020}{\natexlab{b}})}\BibitemShut {NoStop}%
   \bibitem [{\citenamefont {Cao}\ \emph {et~al.}(2018{\natexlab{a}})\citenamefont
     {Cao}, \citenamefont {Fatemi}, \citenamefont {Demir}, \citenamefont {Fang},
     \citenamefont {Tomarken}, \citenamefont {Luo}, \citenamefont
     {Sanchez-Yamagishi}, \citenamefont {Watanabe}, \citenamefont {Taniguchi},
     \citenamefont {Kaxiras}, \citenamefont {Ashoori},\ and\ \citenamefont
     {Jarillo-Herrero}}]{Cao2018}%
     \BibitemOpen
     \bibfield  {author} {\bibinfo {author} {\bibfnamefont {Y.}~\bibnamefont
     {Cao}}, \bibinfo {author} {\bibfnamefont {V.}~\bibnamefont {Fatemi}},
     \bibinfo {author} {\bibfnamefont {A.}~\bibnamefont {Demir}}, \bibinfo
     {author} {\bibfnamefont {S.}~\bibnamefont {Fang}}, \bibinfo {author}
     {\bibfnamefont {S.~L.}\ \bibnamefont {Tomarken}}, \bibinfo {author}
     {\bibfnamefont {J.~Y.}\ \bibnamefont {Luo}}, \bibinfo {author} {\bibfnamefont
     {J.~D.}\ \bibnamefont {Sanchez-Yamagishi}}, \bibinfo {author} {\bibfnamefont
     {K.}~\bibnamefont {Watanabe}}, \bibinfo {author} {\bibfnamefont
     {T.}~\bibnamefont {Taniguchi}}, \bibinfo {author} {\bibfnamefont
     {E.}~\bibnamefont {Kaxiras}}, \bibinfo {author} {\bibfnamefont {R.~C.}\
     \bibnamefont {Ashoori}}, \ and\ \bibinfo {author} {\bibfnamefont
     {P.}~\bibnamefont {Jarillo-Herrero}},\ }\href {\doibase 10.1038/nature26154}
     {\bibfield  {journal} {\bibinfo  {journal} {Nature}\ }\textbf {\bibinfo
     {volume} {556}},\ \bibinfo {pages} {80–84} (\bibinfo {year}
     {2018}{\natexlab{a}})}\BibitemShut {NoStop}%
   \bibitem [{\citenamefont {Cao}\ \emph {et~al.}(2018{\natexlab{b}})\citenamefont
     {Cao}, \citenamefont {Fatemi}, \citenamefont {Fang}, \citenamefont
     {Watanabe}, \citenamefont {Taniguchi}, \citenamefont {Kaxiras},\ and\
     \citenamefont {Jarillo-Herrero}}]{Cao2018a}%
     \BibitemOpen
     \bibfield  {author} {\bibinfo {author} {\bibfnamefont {Y.}~\bibnamefont
     {Cao}}, \bibinfo {author} {\bibfnamefont {V.}~\bibnamefont {Fatemi}},
     \bibinfo {author} {\bibfnamefont {S.}~\bibnamefont {Fang}}, \bibinfo {author}
     {\bibfnamefont {K.}~\bibnamefont {Watanabe}}, \bibinfo {author}
     {\bibfnamefont {T.}~\bibnamefont {Taniguchi}}, \bibinfo {author}
     {\bibfnamefont {E.}~\bibnamefont {Kaxiras}}, \ and\ \bibinfo {author}
     {\bibfnamefont {P.}~\bibnamefont {Jarillo-Herrero}},\ }\href {\doibase
     10.1038/nature26160} {\bibfield  {journal} {\bibinfo  {journal} {Nature}\
     }\textbf {\bibinfo {volume} {556}},\ \bibinfo {pages} {43–50} (\bibinfo
     {year} {2018}{\natexlab{b}})}\BibitemShut {NoStop}%
   \end{thebibliography}


\end{document}


\title{Supplemental information for: Anomalous photo-induced band renormalization in correlated materials: Case study of Ta$_2$NiSe$_5$}

\author{Lei Geng}
\author{Xiulan Liu}
\author{Jianing Zhang}
\author{Denis Gole{\v{z}}}
\author{Liang-You Peng}

\maketitle

\section{A. Model built from {\it ab initio} calculations}

To develop a model that faithfully describes the real material, we conducted {\it ab initio} calculations for TNS and extracted essential parameters. We followed the methodology used in Ref.~\cite{mazza2020nature}. The first step is to obtain the energy bands through density functional theory~(DFT) calculations. Our calculations utilized the Vienna ab initio simulation package (VASP)~\cite{kresse1996efficient}. In reality, the two-dimensional material TNS comprises two distinct layers within a unit cell. However, for the sake of simplicity, we adopted a unit cell containing 16 atoms within the same layer, as a convention in previous work~\cite{windgatter2021common}. TNS exhibits two structure phases: the orthorhombic phase and the monoclinic phase. The former is a semimetal phase existing in the high temperature, while the latter is an insulator phase existing in the low temperature. We want to reproduce the spontaneous symmetry breaking from the orthorhombic phase to the monoclinic phase through pure electronic mechanisms~\cite{mazza2020nature}. Therefore, we extract the parameters from the orthorhombic phase to build the model. Simultaneously, it is essential to compare the energy band results in the monoclinic phase with the symmetry-breaking outcomes of our model. Thus, the calculations for both phases are needed. However, the simulation for the monoclinic phase within the framework of DFT presents significant challenges. As stated in Ref.~\cite{windgatter2021common}, we employed the results from a triclinic cell to approximate those of the monoclinic cell. The material is also an insulator with this structure. And the corresponding energy band of the insulator phase is shown in Fig.~1(a) in the main text. Here, our primary focus will be on the DFT results pertaining to the orthorhombic phase. In the calculations, we employed the lattice parameter from Ref.~\cite{subedi2020orthorhombic} for the orthorhombic phase and utilized the modified Becke-Johnson (mBJ) functional with the parameter $c_{mbj}=1.259$. The first Brillouin zone was discretized using a $20\times20\times3$ k-grid.

\begin{figure}[!ht]
  \centering
  \begin{subfigure}{0.45\linewidth}
    \parbox[][8cm][c]{\linewidth}{
      \centering
      \includegraphics[scale=0.5]{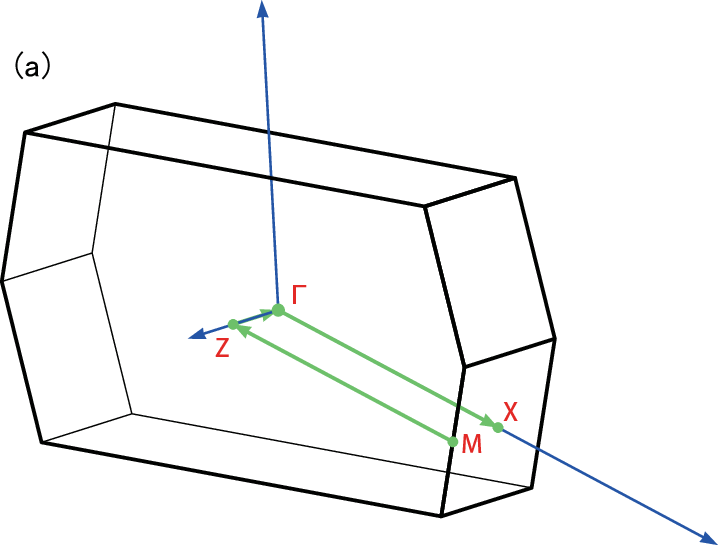}
    }
  \end{subfigure}
  \begin{subfigure}{0.45\linewidth}
    \parbox[][8cm][c]{\linewidth}{
      \centering
      \includegraphics[scale=0.15]{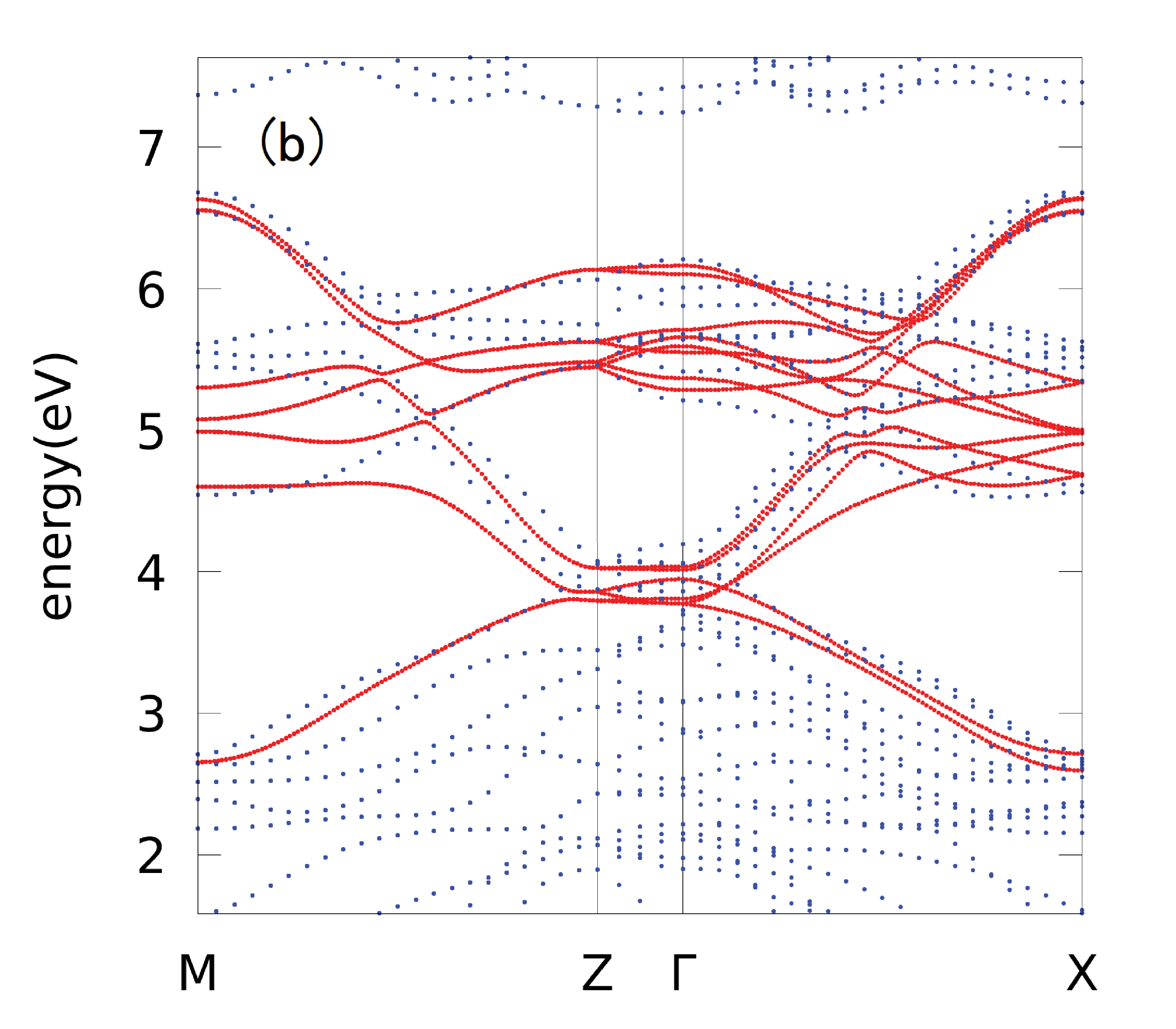}
    }
  \end{subfigure}
  \captionsetup{justification=raggedright}
  \caption{(a) The high-symmetry points in the first Brillouin zone of TNS. (b) The blue dots represent energy bands of DFT results and the red dots represent those of MLWFs.}
  \label{SM_fig1}
\end{figure}

The second step involves computing maximally localized Wannier functions~(MLWFs) based on the DFT results. For this purpose, we utilized the Wannier90 package~\cite{mostofi2008wannier90}. In order to avoid artificial effects, the frozen window around the Fermi surface was not used. In total, we acquired 14 MLWFs in a unit cell, with 12 orbitals centered on \ce{Ta} atoms and 2 orbitals centered on \ce{Ni} atoms. In Fig.~\ref{SM_fig1}(b), we compare the energy bands of the MLWFs with the original DFT results. It is evident that they match closely, indicating that the MLWFs provide an accurate description of this system. 

\begin{figure}[htb]
  \centering
   \includegraphics[width=\linewidth]{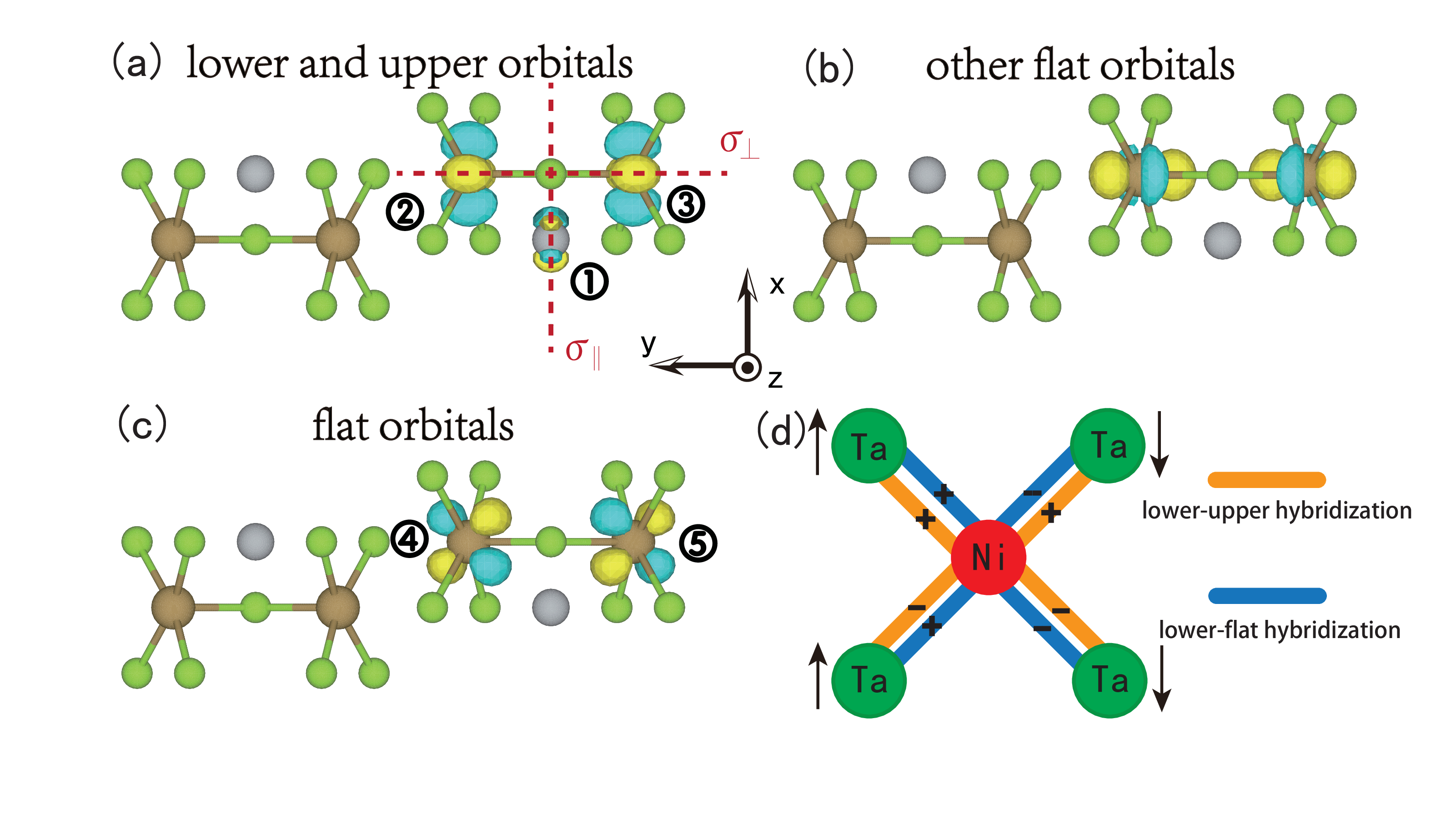}
  \captionsetup{justification=raggedright}
  \caption{(a) Two upper orbitals and one lower orbital are shown with the symmetry plane $\sigma_{\parallel}$ and $\sigma_{\perp}$. (b) The flat orbitals that are not considered in our model. (c) The flat orbitals considered in our model. (d) A sketch of the hybridization for the nearest-neighbor \ce{Ta}-\ce{Ni} atoms. The orange lines represent the hybridization between the lower and the upper orbitals and the blue lines represent that between the lower and the flat orbitals.
  }
  \label{SM_fig2}
\end{figure}

It's necessary to analyze the symmetry of all the orbitals. The inter-chain interactions have been disregarded. Hence, we have focused on seven orbitals in a single chain to construct our model. The shapes of these orbitals are depicted in Fig.~\ref{SM_fig2}. The lower orbital denoted by \textcircled{1} is symmetric about the plane $\sigma_{\parallel}$ but asymmetric about the plane $\sigma_{\perp}$. Upper orbitals denoted by \textcircled{2} and \textcircled{3} are symmetry about both the plane $\sigma_{\parallel}$ and $\sigma_{\perp}$. The results of these orbitals are close to those presented in Ref.~\cite{mazza2020nature}. Moreover, our model incorporates flat orbitals with higher energies, denoted as \textcircled{4} and \textcircled{5}. These orbitals exhibit asymmetry with respect to both the $\sigma_{\parallel}$ and $\sigma_{\perp}$ plane. In fact, there are additional flat orbitals contributing to the flat band, which we have depicted in Fig.~\ref{SM_fig2}(b) without numeric labels. These orbitals have the same symmetry as upper orbitals. However, they were not included in our calculations due to the limit of the memory. And they are not as important as the flat orbitals \textcircled{4} and \textcircled{5}. We will discuss the details in the following. The symmetry of all these orbitals will determine the hybridizations (hopping of different orbitals) between them. Hybridizations of orbitals 1-5 on the nearest-neighbor \ce{Ta}-\ce{Ni} atoms are illustrated in Fig.~\ref{SM_fig2}(d). Specifically, for a lower orbital on the \ce{Ta} atom, its hybridization with upper orbitals on the nearest-neighbor \ce{Ta} atom exhibits symmetry about $\sigma_{\parallel}$ but asymmetry about $\sigma_{\perp}$. In contrast, its hybridization with flat orbitals on these atoms exhibits asymmetry about $\sigma_{\parallel}$ but symmetry about $\sigma_{\perp}$. When these hybridizations are considered collectively, they yield the monoclinic structural symmetry. This explains why the material undergoes a monoclinic phase transition rather than a triclinic phase transition. At higher temperatures, due to the much lower energy of the upper band bottom, the population of the upper band exceeds that of the flat band. As a result, the structural symmetry is predominantly influenced by the symmetries of lower and upper orbitals, leading to the orthorhombic system. Conversely, at relatively low temperatures, the populations of upper and flat orbitals become comparable. This occurs because the hybridization between lower and flat orbitals is significantly stronger than that between lower and flat orbitals. We will elaborate on this point by hopping parameters between various Wannier orbitals in the following. Under this condition, the structural symmetry is influenced by both hybridization symmetries, resulting in the monoclinic system. 

In addition to symmetry considerations, we obtained specific hopping parameters from the Wannnier90, which can be utilized in the minimal model. Within the same cell, the hopping parameters are as follows: $t_{lu}=0.078~{\rm eV}$ between the lower and the upper orbitals, $t_{lf}=0.312~{\rm eV}$ between the lower and the flat orbitals, and the hopping between the upper and the flat orbitals on the same site is zero, as dictated by symmetry. Furthermore, the hybridization between the lower orbitals and the omitted flat orbitals is $t_{lf2}=0.213~{\rm eV}$. Regarding the hopping of the same orbitals in the nearest-neighbor cell along the $x$ direction, the values are as follows: $t_{ll}=0.320~{\rm eV}$ for lower orbitals, $t_{uu}=-0.467~{\rm eV}$ for upper orbitals and $t_{ff}=0.123~{\rm eV}$ for flat orbitals. These parameters were moderately adjusted to fit the spectral function more closely with the DFT energy band. The obtained single-particle part of the Hamiltonian $\epsilon(k)$ is written as
\[
\epsilon(k) =0.3~{\rm eV}\times \begin{bmatrix}
2\cos(k)-4.18 & 0.25*[1-\exp(ik)]  & -0.25*[1-\exp(-ik)] & 0.8*[1+\exp(ik)] & -0.8*[1+\exp(-ik)] \\
0.25*[1-\exp(-ik)]     & 0.34-4.4\cos(k)               & 0               & 0 & 0 \\
-0.25*[1-\exp(ik)]    & 0    & 0.34-4.4\cos(k)        & 0               & 0 \\
0.8*[1+\exp(-ik)]      &0     & 0           & 2.32             & 0 \\
-0.8*[1+\exp(ik)]      &0     & 0           & 0               &2.32
\end{bmatrix}.  
\]
In the matrix, we use the length of the unit cell in the $\Gamma\text{-}X$ direction as the unit length, and the wave vector $k$ also aligns with the $\Gamma\text{-}X$ direction. Due to the symmetry, the hybridization between the lower orbital and upper orbitals is zero at the $\Gamma$ point, as well as that between the lower orbital and omitted flat orbitals. Because the pump laser is resonant for the transition from the valence top to flat bands. The hybridization near the $\Gamma$ point will determine the transition probability when the pump laser is applied. Therefore, we believe that electrons in the lower orbitals are more likely to be excited to the flat orbitals \textcircled{4} and \textcircled{5}, rather than the omitted flat orbitals. That is one of the reasons why we have excluded those flat orbitals from the model. 

In addition to hopping parameters, the position matrix is also a crucial parameter used in the length gauge of the Hamiltonian. We mentioned that the approximation $\left<0\alpha|r|R\alpha'\right>=\delta_{0R}\delta_{\alpha\alpha'}\tau_{\alpha}$ is used in the main text. To demonstrate the validity of this approximation, we should compare position matrix elements for the same orbital and different orbitals. As examples, we obtain some elements in $y$ direction, $\left<04|y|04\right>=2.06$~\AA, $\left<01|y|01\right>=0$~\AA, $\left<04|y|01\right>=-0.085$~\AA~and $\left<02|y|01\right>=-0.041$~\AA. Due to the localization of MLWFs, it is evident that the diagonal elements of the position operator are much larger than the off-diagonal terms.

The third step is to obtain the many-body density-density interactions for these orbitals, denoted as term $H_U$ and $H_V$ in the main text. For implementation, we employed the constrained random phase approximation~(cRPA) method~\cite{aryasetiawan2004frequency} in VASP. Our results closely resemble those reported in Ref.~\cite{mazza2020nature}. We obtained the Hubbard interactions $U_{\rm Ta}=2.1~{\rm eV}$ and $U_{\rm Ni}=2.5~{\rm eV}$. The interaction between different orbitals on the nearest-neighbor \ce{Ta}-\ce{Ni} atoms is $V_{\rm TaNi}=1.0~{\rm eV}$. The distinction between flat orbitals and upper orbitals is negligible here; both can be effectively considered as \ce{Ta} orbitals. In addition, the interaction between flat and upper orbitals on the same atom is represented as $V_{\rm Ta}=1.6~{\rm eV}$. It is worth stressing that we did not consider the spin-orbital coupling in this study. Spin-up and spin-down orbitals are degenerate, hence, only one branch is considered in our model. We maintain the ratio of interaction strengths above and apply a scaling factor. The interaction we used in the reciprocal space on the basis of five orbitals can be written as
\[
V(k) =0.75~{\rm eV}\times \begin{bmatrix}
1.25 & 1+\exp(ik) & 1+\exp(-ik) & 1+\exp(ik) & 1+\exp(-ik) \\
1+\exp(-ik)     & 1.05               & 0               & 1.6 & 0 \\
1+\exp(ik)      & 0    & 1.05        & 0               & 1.6 \\
1+\exp(-ik)     &1.6     & 0         & 1.05            & 0 \\
1+\exp(ik)      &0     & 1.6         & 0               & 1.05
\end{bmatrix}.  
\]
The diagonal terms have been scaled by a factor of 0.5 because the Hubbard interaction only arises between orbitals with different spins. We apply the GW approximation to handle the interaction in our study. In principle, the Hubbard interaction $H_{\rm U}$ should be dealt within the SGW approximation~\cite{von2010kadanoff}. However, the higher-order self-energy terms are not expected to significantly impart the result in this study. Therefore, we only omit the Fock term of the Hubbard interaction and handle it using the GW approximation for the sake of simplicity. Regarding the interaction $H_{\rm V}$, since it is spin-independent, we do not make assumptions about whether the exciton is in a singlet or triplet state. Both singlet and triplet states are permitted in our model. In addition, we only consider the interaction between orbitals on the same atom or the nearest-neighbor atoms in this model. If we consider the interaction between orbitals on atoms in a larger distance like two nearest \ce{Ta} atoms. The Hartree shift of Eq.~(5) in the main text may change. According to the parameters in our calculations, the values of the Hartree shift will be close to zero. However, this is the consequence of the fact that the interaction does not decrease as fast as $1/r$, which may be attributed to the inaccuracy of the method for the long-range interaction. We think it is very likely that the reality is close to our nearest-neighbour model because of the consistency of our simulation results and the experimental observations.

\begin{figure}
  \includegraphics[width=0.8\linewidth]{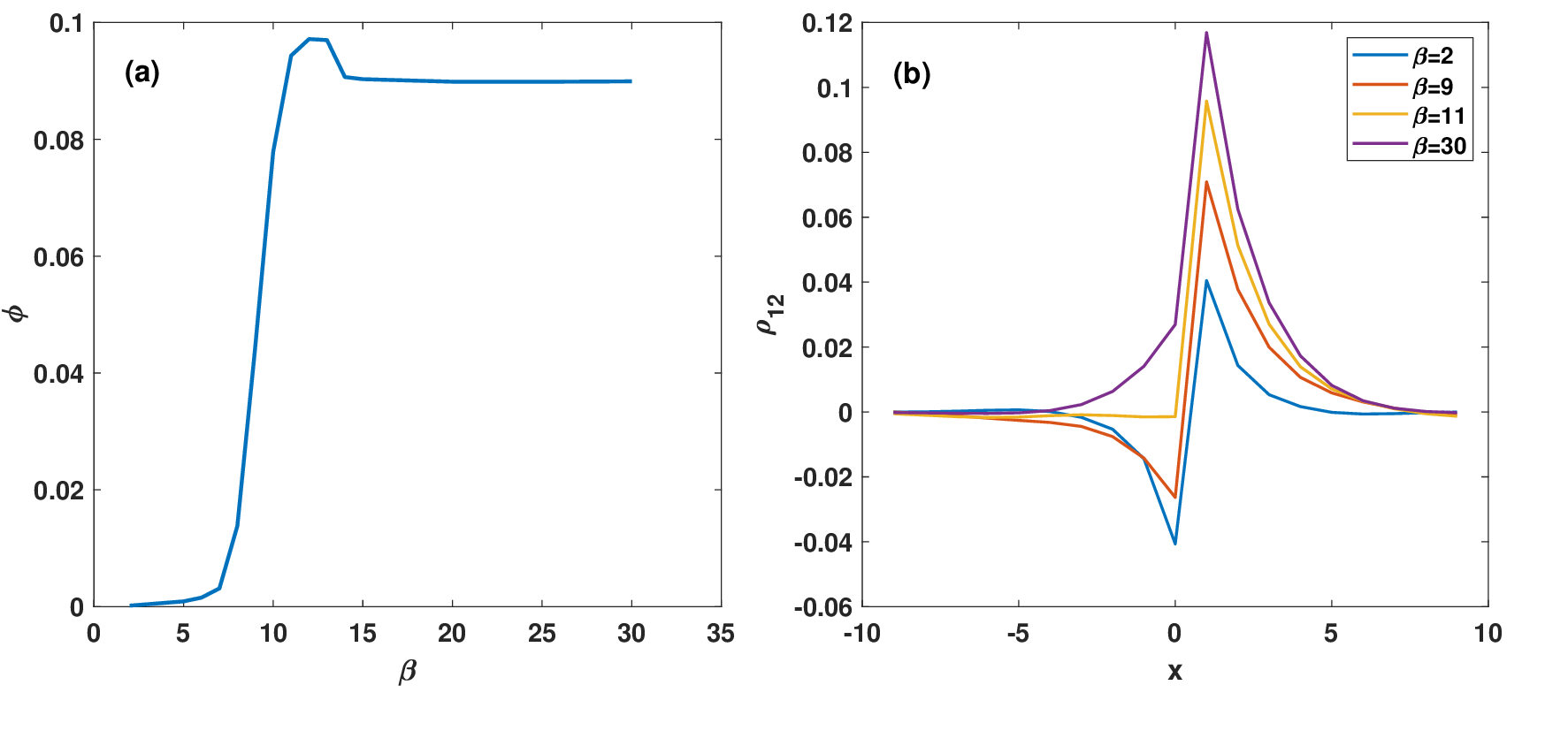}\quad
  \includegraphics[width=\linewidth]{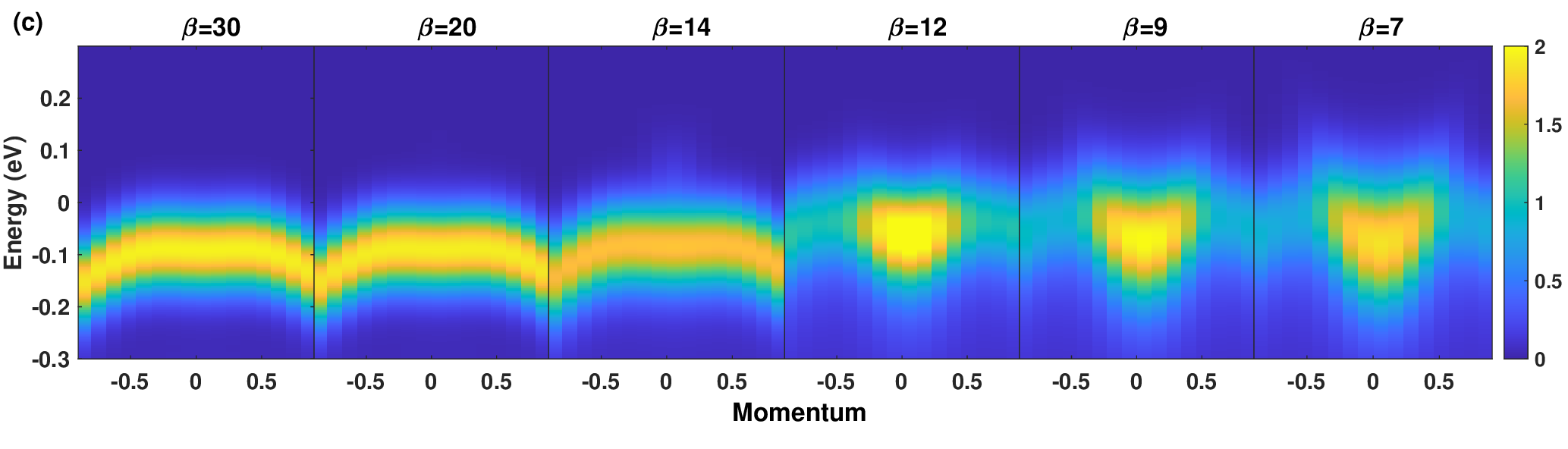}
  \captionsetup{justification=raggedright}
  \caption{(a) The inverse temperature dependence of the order parameter $\phi$. (b) The inverse temperature dependence of the element of density matrix $\rho_{12}(x)$. (c) The simulated photoemission spectrum of the equilibrium state near the Fermi surface for different inverse temperatures.
  }
  \label{SM_fig3}
\end{figure}

The final step involves incorporating the previously parameters mentioned above into the minimal model and solving the non-equilibrium Green's function in the time domain. Prior to investigating the laser-induced non-equilibrium problem, it's essential to examine the equilibrium characteristics of this model. The unit energy is set to 0.3eV, we use 56 k-grids and a time step $dt=0.02$ in the equilibrium and non-equilibrium simulations. The spectral function of this model at a low temperature $\beta=30~(T=116~{\rm K})$ has been shown in Fig.~1(a) in the main text. An apparent energy gap of appropriately $0.2~{\rm eV}$ can be observed, closely resembling the experimental value of $0.16~{\rm eV}$~\cite{windgatter2021common}. As the temperature increases, the system undergoes an insulator-semimetal phase transition, accompanied by a significant decrease in the order parameter $\phi$ as defined in the main text. Fig.~\ref{SM_fig3}(a) illustrates the plot of the order parameter against the inverse temperature $\beta$. In Fig.~\ref{SM_fig3}(b), we present the distributions of the elements of the density matrix $\rho_{12}$ in real space at various inverse temperatures. The phase transition point is noticeable at around the inverse temperature $\beta=10$, corresponding to a temperature of $T=348~{\rm K}$. This is also close to the experimental value $T_c=326~{\rm K}$. Fig.~\ref{SM_fig3}(a) reveals a slight increase before the decrease of the order parameter as the temperature rises. This occurs because the element of the density matrix $\rho_{12}(x)$ switches from being symmetric about $x=1$ to being asymmetric about $x=0.5$ during this process. Fig.~\ref{SM_fig3}(c) illustrates the variation in the simulated photoemission spectrum at various inverse temperatures. Below the inverse temperature $\beta=14$, it remains nearly unchanged except for a minor elevation of the valence band. A sharp transition occurs between $\beta=14$ and $\beta=12$. The `M`-shaped valence band top disappears, and the system gradually undergoes a transition into a semimetal phase. An evident lift for the top occurs during this process. As the temperature continues to increase, the spectrum becomes blurred. At the same time, the lower edge of the valence top may become even lower than that in the insulator phase. All these characteristics are similar to the experimental findings in Ref.~\cite{tang2020non}, which further validates the reliability of our model.

The Hamiltonian of the model will become Eq.~(5) in the main text when one considers the light-matter interaction. For the case in this work, the frequency of the pump pulse is much higher than the band gap, rendering the drift $k-A$ negligible and thus ignorable. The term $E\cdot\tau_{\alpha}$ will govern the laser-induced dynamics. Its value varies depending on the polarization of the pump pulse. Since TNS is a two-dimensional layered material, its value is notably larger in $x$ and $y$ directions. We set the center of $\ce{Ni}$ as the origin, which leads to $\tau_2=-\tau_3=\tau_4=-\tau_5$ in these two directions. The laser-induced vibrating term $E\cdot\tau_{\alpha}$ is diagonal rather than off-diagonal on the basis of Wannier orbitals. So the transfer between two orbitals is not only related to the laser intensity but also depends on the off-diagonal hybridization terms in the Hamiltonian. The large hybridization between the lower orbitals and the flat orbitals in the matrix $\epsilon(k)$ implies the strong excitation in the flat bands.

\section{B. Parameter dependence of lifetime of the long-lived state}

\begin{figure}[htb]
  \includegraphics[width=0.8\linewidth]{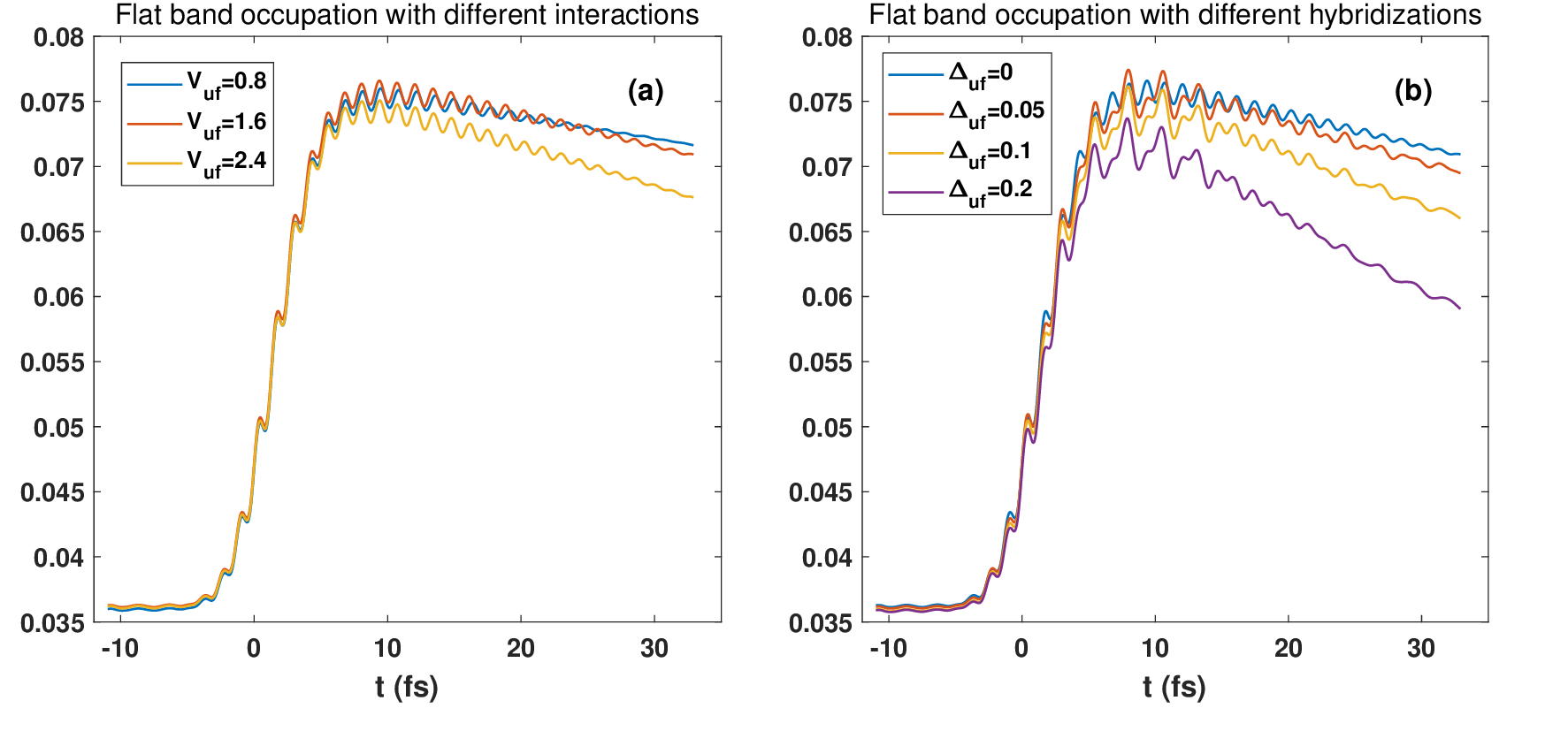}
  \captionsetup{justification=raggedright}
  \caption{(a) The flat band occupation $\frac{\rho_{44}+\rho_{55}}{2}$ for different interactions $V_{uf}$ between the upper orbitals and the flat orbitals. (b) The flat band occupation $\frac{\rho_{44}+\rho_{55}}{2}$ for different hybridizations $\Delta_{uf}$ between the upper orbitals and the flat orbitals.
  }
  \label{SM_fig4}
\end{figure}

In the main text, we demonstrated the long lifetime of the nonthermal distribution in flat bands and attribute it to the absence of hybridization between the flat and the upper orbitals, arising from their distinct parities in the orthorhombic symmetry. However, this symmetry undergoes a slight breaking in the monoclinic phase. To investigate the robustness of long-lived state and the model's parameter dependence, we varied the hybridization and interaction between upper orbitals and flat orbitals as $\epsilon_{24}=\epsilon_{42}=\epsilon_{53}=\epsilon_{35}=0.3~{\rm eV}\times\Delta_{uf}$ and $V_{24}=V_{42}=V_{35}=V_{53}=0.75~{\rm eV}\times V_{uf}$. The results are depicted in Figs.~\ref{SM_fig4}(a) and (b). Remarkably, we can observe the robustness of the long lifetime for small variations in the hybridization. Decreased hybridizations and interactions lead to a more constrained pathway to the prethermalization, indicating a prolonged lifetime.

\section{C. Calculation of the spectral function and the photoemission spectrum}

In this study, two important quantities derived from our model include the spectral function and the photoemission spectrum. The spectral function is actually the momentum-resolved density of the state. This allows for a qualitative comparison with the energy band obtained from DFT, the difference is that our spectral function exhibits a finite width due to the non-equal-time self-energy. The photoemission spectrum is what is actually measured in the tr-ARPES experiment, which corresponds to the energy-resolved and momentum-resolved electron occupation in the material. In the experiment, photoelectrons are excited by a weak, high-frequency probe laser and captured by the detector. Whereas we can calculate the electron distribution directly without incorporating the probe pulse in our program. Both the spectral function and the photoemission spectrum are computed through the Fourier transformation of the Green's function in the time domain. The spectral function corresponds to the retarded Green's function $G^{\rm ret}_{\alpha\alpha,k}(t,t')=-i\theta(t-t')\left< \{ a_{\alpha,k}(t),a^{\dagger}_{\alpha,k}(t') \} \right>$, and the photoemission spectrum corresponds to the lesser Green's function $G^{<}_{\alpha\alpha,k}(t,t')=i\left<a^{\dagger}_{\alpha,k}(t')a_{\alpha,k}(t)\right>$. In principle, we need to do a two-dimensional Fourier transformation for two time variables $t$ and $t'$. Taking the photoemission spectrum as an example, one has
\begin{equation}
  I(\omega,k,t_p)=-{\rm Im}\left[\sum_{\alpha}\int_{-\infty}^{\infty}\int_{-\infty}^{\infty}dt dt'S(t'-t_p)S(t-t_p)\exp(i\omega(t'-t))G^{<}_{\alpha\alpha,k}(t,t')\right],
\label{PhotoemissionSpectrum}
\end{equation}
where $S(t)=\exp(-t^2/\delta^2)$ is the envelope of the probe pulse. However, we only perform a single integral to substitute for simplicity
\begin{equation}
  I(\omega,k,t_p)=-{\rm Im}\left[\sum_{\alpha}\int_{-\infty}^{t_p}dt' S(t'-t_p)\exp(i\omega(t'-t_p))G^{<}_{\alpha\alpha,k}(t_p,t')\right].
\label{PhotoemissionSpectrumSimple}
\end{equation}
Its result is similar to the double integral.

Another point that needs to be emphasized is that the observed photoemission spectra in experiments cannot be exactly equal to the numerical results obtained by summing over all the bands in principle. In a typical experiment, the scattering plane usually coincides with the $\sigma_{\perp}$ plane, and the probe pulse typically exhibits even or odd parity about this plane~\cite{watson2020band}. Consequently, it can only detect electronic states with even or odd parity about the plane $\sigma_{\perp}$. Since most experiments concentrate on the electronic structure near the $\Gamma$ point, and upper orbitals (which have even parity about the $\sigma_{\perp}$ plane in the orthorhombic phase) contribute the most in this region. The polarization of the probe pulse is typically in the plane $\sigma_{\perp}$. While it may seem that summing over the index $\alpha$ for the upper orbitals would be sufficient to compare with experimental results. However, it is important to note that the plane $\sigma_{\perp}$ is no longer a strictly symmetric plane in the monoclinic phase. We have checked that the behavior near the $\Gamma$ point of the result summed for all the bands is very similar to that with only upper bands. Therefore, in the main text, we present spectral functions and photoemission spectra for all the bands. This approximation does not qualitatively change our conclusions of the present work.